\documentclass[manuscript]{aastex}
\usepackage{natbib}

\usepackage{color}

\slugcomment{Resubmitted to ApJ}

\shorttitle{Absorption and Reddening in Mrk 231}
\shortauthors{Leighly et al.}

\begin{document}


\title{Evidence for AGN Feedback in the Broad Absorption Lines and
  Reddening of Mrk 231\footnote{This work is based on
    observations obtained at the MDM Observatory, operated by
    Dartmouth College, Columbia University, Ohio State University,
    Ohio University, and the University of
    Michigan.}$^{\, ,\,}$\footnote{Based on observations obtained with 
    the Apache Point Observatory 3.5-meter telescope, which is owned
    and operated by the Astrophysical Research Consortium.}}


\author{Karen M. Leighly\footnote{Visiting Astronomer at the Infrared
    Telescope Facility, which is operated by the University of Hawaii
    under Cooperative Agreement no. NNX-08AE38A with the National
    Aeronautics and Space Administration, Science Mission
    Directorate, Planetary Astronomy Program.}$^{\, ,\, }$\footnote{Visiting
    Astronomer, Kitt Peak National Observatory, which is operated by
    the Association of Universities for Research in Astronomy (AURA)
    under cooperative agreement with the National Science
    Foundation.}}
\affil{Homer L.\ Dodge Department of Physics and Astronomy, The
  University of Oklahoma, 440 W.\ Brooks St., Norman, OK 73019}

\author{Donald M. Terndrup$^3$}
\affil{Department of Astronomy, The Ohio State University, 140
  W.\ 18th Ave., Columbus, OH 43210} 

\author{Eddie Baron}
\affil{Homer L.\ Dodge Department of Physics and Astronomy, The
  University of Oklahoma, 440 W.\ Brooks St., Norman, OK 73019}

\author{Adrian B.\ Lucy}
\affil{Homer L.\ Dodge Department of Physics and Astronomy, The
  University of Oklahoma, 440 W.\ Brooks St., Norman, OK 73019}

\author{Matthias Dietrich$^{3,4}$}
\affil{Department of Physics and Astronomy, Ohio University,
  Clippinger Labs 251B, Athens, OH 45701}

\and

\author{Sarah C. Gallagher}
\affil{Department of Physics \& Astronomy, The University of Western
  Ontario, London, ON, N6A 3K7, Canada}




\begin{abstract}
We present the first J-band spectrum of Mrk 231, which reveals a large
\ion{He}{1}*$\lambda 10830$ broad absorption line with a profile
similar to that of the well-known \ion{Na}{1} broad absorption line.
Combining this spectrum with optical and UV spectra from the
literature, we show that the unusual reddening noted by
\citet{veilleux13} is explained by a reddening curve like those
previously used to explain low values of total-to-selective extinction
in SNe Ia.  The nuclear starburst may be the origin and location of
the dust.   { Spatially-resolved emission in the broad absorption 
line trough suggests nearly full coverage of the continuum emission
region.}  The broad absorption lines reveal higher velocities in the
\ion{He}{1}* lines (produced in the quasar-photoionized \ion{H}{2}
region) compared with the \ion{Na}{1} and \ion{Ca}{2} lines (produced
in the corresponding partially-ionized zone).  {\it Cloudy}
simulations show that a density increase is required between the
\ion{H}{2} and partially-ionized zones  to produce ionic column
densities consistent with the optical and IR absorption line
measurements and limits, and that the absorber lies $\sim 100\rm \,
pc$ from the central engine. These results suggest that the
\ion{He}{1}* lines are produced in an ordinary quasar BAL wind that
impacts upon, compresses, and accelerates the nuclear starburst's
dusty effluent (feedback in action), and the \ion{Ca}{2} and
\ion{Na}{1} lines are produced in this dusty accelerated gas. This
unusual circumstance explains the rarity of \ion{Na}{1} absorption
lines; without the compression along our line of sight, Mrk~231 would
appear as an ordinary FeLoBAL.

\end{abstract}


\section{Introduction\label{intro}}

Mrk~231 is a nearby ($z=0.0421$) ultraluminous infrared galaxy that
has a Seyfert 1 optical spectrum \citep{sanders88}. The infrared
emission is thought to be a combination of AGN and starburst activity
\citep[e.g.,][and references therein]{farrah03}.  Recently, attention
has been again drawn to this galaxy as a consequence of the discovery
of a powerful, wide-angle, kiloparsec-scale molecular outflow
\citep{rupke11}.   

Mrk 231's optical spectrum shows extreme \ion{Fe}{2} emission,
undetected [\ion{O}{3}]$\lambda\lambda 4959,5007$, and somewhat broad
and prominent Balmer 
lines.  The most remarkable feature is the very strong broad
\ion{Na}{1}D absorption line with  $v \approx -4,500 \rm \, km\,
s^{-1}$  \citep{aw72,boksenberg77, rudy85, sm85, hk87, boroson91,
  kdh92,   forster95,rupke02, gallagher05, lipari05,  rupke05,
  rodriguez09,   veilleux13}, not to be confused with the few-hundred
$\rm km\, s^{-1}$ \ion{Na}{1}D absorption consistent with the velocity
of the molecular outflow \citep{rupke05}.  While low-velocity \ion{Na}{1}D
absorption from molecular outflows is relatively common
\citep[e.g.,][]{rupke05}, high-velocity, broad absorption \ion{Na}{1}D
lines are rare. The \ion{Na}{1}D absorption is accompanied by
\ion{Ca}{2}$\lambda 3935,3970$ and \ion{He}{1}*$\lambda 3889$  in the
optical bandpass \citep[e.g.,][]{aw72,boksenberg77,rudy85,sm85,hn87,
  boroson91,rupke02,lipari05,rodriguez09, veilleux13}.  Near-UV and UV
broad absorption lines were reported by \citet{smith95} and
\citet{gallagher05}.  

{ The broad absorption lines classify
  Mrk~231 as a broad absorption line Quasar (BALQSO).  Broad
  absorption line quasars constitute between 10 and 40\% of quasars,
  depending on selection criteria \citep[e.g.,][]{hw03, trump06, dai08}.}
Mrk~231 also exhibits  \ion{Fe}{2} broad absorption lines, making
Mrk~231 the nearest iron low-ionization broad absorption line quasar
(FeLoBAL). { FeLoBALs are much rarer than BALQSOs, constituting
  only $\sim 2$\% of quasars, and again, the rate of incidence
  measured depends on the sample selection \citep{urrutia09,
    dai12}. BALQSOs have drawn intense interest in recent years, as
  their outflows may be key for understanding AGN feedback, i.e., the
  process by which an AGN influences its host galaxy, specifically,
  how does the outflow act to supress specifically, how it acts to
  suppress the rate of star formation in the host; see, for example, 
\citet{farrah12} who provides a discusison of the problem, and also 
has recently found evidence for an anticorrelation
between the strength of the broad absorption-line outflow and the
fractional contribution from star formation to the IR luminosity in a
sample of FeLoBALs; they interpret this result as evidence for the AGN
outflow curtailing star formation in the host galaxy.}  

As noted by \citet{veilleux13}, the presence of \ion{He}{1}*$\lambda
3889$ and \ion{Na}{1} in the same outflow appears to be problematic
from a photoionization point of view.   Metastable \ion{He}{1}*
is formed by recombination onto He$^+$, and so 
it occurs in the \ion{H}{2} region of the outflow, in roughly the same
gas that would generate \ion{C}{4} \citep[e.g.,][]{leighly11}.  In
contrast, neutral sodium has  an ionization potential of only $5.14\rm
\, eV$, i.e., less than that of hydrogen, so must occur in the
partially-ionized zone beyond the hydrogen ionization front.  It is
distinguished from other low-ionization absorption lines such as
\ion{Fe}{2} that are formed in the partially-ionized zone, however.
As shown in \citet{lucy14}, copious \ion{Fe}{2} is produced just past
the hydrogen ionization front; \ion{Na}{1}D is produced in basically
neutral gas, much deeper in the slab, and farther from the illuminated
face. Thus, \ion{He}{1}* and \ion{Na}{1} cannot be produced in gas
with even close to the same ionization state, and  so it is not
necessarily easy to see how they can exhibit the same  dynamics.   

Mrk~231 also has an unusual optical/UV continuum spectrum
\citep{smith95,   veilleux13}.  Specifically, while the optical
spectrum shows only modest reddening, the spectrum falls steeply in
the near UV, and levels out shortward of $\sim 2400$\AA\/.  Anomalously steep
reddening has been seen in several BALQSOs \citep{hall02,  leighly09,
  jiang13}, but the mechanism for this reddening remains unexplained.   

In the paper, we present new infrared spectra from Mrk~231. While
Mrk~231 has been observed often in the 
infrared, a careful search of the literature revealed no previously
published J-band spectra. Our spectrum reveals, for the first time, the
broad \ion{He}{1}*$\lambda 10830$ absorption line 
(\S\ref{irtf}).   We also find evidence for the appearance
of a new \ion{He}{1}*$\lambda 10830$ velocity component at $11,520 \rm
km \, s^{-1}$ (\S\ref{mdm}, \S\ref{apo}).  We present new, high
signal-to-noise-ratio 
blue optical spectra (\S\ref{kpno}).  Combining our data with
published spectra (\S\ref{digitize}),   we present an analysis of the
reddening in this object, finding that the unusual spectral shape is
consistent with so-called ``circumstellar reddening'' (to be defined
and discussed in \S\ref{goobar}).  \S\ref{abs_lines}  presents an
analysis of the spectra, and extraction of the \ion{He}{1}* and
\ion{Ca}{2} line profiles, which we compare with a \ion{Na}{1} profile
obtained from the literature.  After introducing a physical scenario
for the absorption lines in \S\ref{picture}, we present in
\S\ref{sims} a simple  photoionization model, using {\it   Cloudy},
that yields the ionic column densities required to produce the
observed absorption lines. We systematically investigate the
assumptions of our model, within the limitations of {\it Cloudy}
in \S\ref{beyond} and provide an optimal model in \S\ref{optimize}.
The results are summarized in \S\ref{summary}.   

\begin{deluxetable}{ccc}
\scriptsize
\tablewidth{0pt}
\tablecaption{IRTF Spectrum}
\tablehead{
 \colhead{Observed Wavelength} & \colhead{Flux Density\tablenotemark{a}} &
 \colhead{Flux Density Error\tablenotemark{a}} \\
 \colhead{($\mu$m)} & \colhead{($10^{-17}\rm \,
   erg\, s^{-1}\, cm^{-2}\, $\AA\/$^{-1}$)} & \colhead{($10^{-17}\rm \,
   erg\, s^{-1}\, cm^{-2}\, $\AA\/$^{-1}$)}}
\startdata
      0.805413    &   988.822 &      10.4281 \\
      0.805616    &   994.342 &      10.8305 \\
      0.805818    &   988.422 &      10.5220 \\
      0.806020    &   987.099 &      10.8525 \\
      0.806223    &   971.811 &      10.6738 \\
\enddata
\tablecomments{Table \ref{irtf_spec_table} is published in its entirety in the
  electronic edition of the {\it Astrophysical Journal}.  A portion is
  shown here for guidance regarding its form and content.}
\tablenotetext{a}{Corrected for Galactic extinction.}
\label{irtf_spec_table}
\end{deluxetable}
\normalsize

\section{Observations and Data\label{obs}}

\subsection{IRTF SpeX Observation\label{irtf}}

\begin{figure}[!h]
\epsscale{1.0}
\begin{center}
\includegraphics[width=4.0truein]{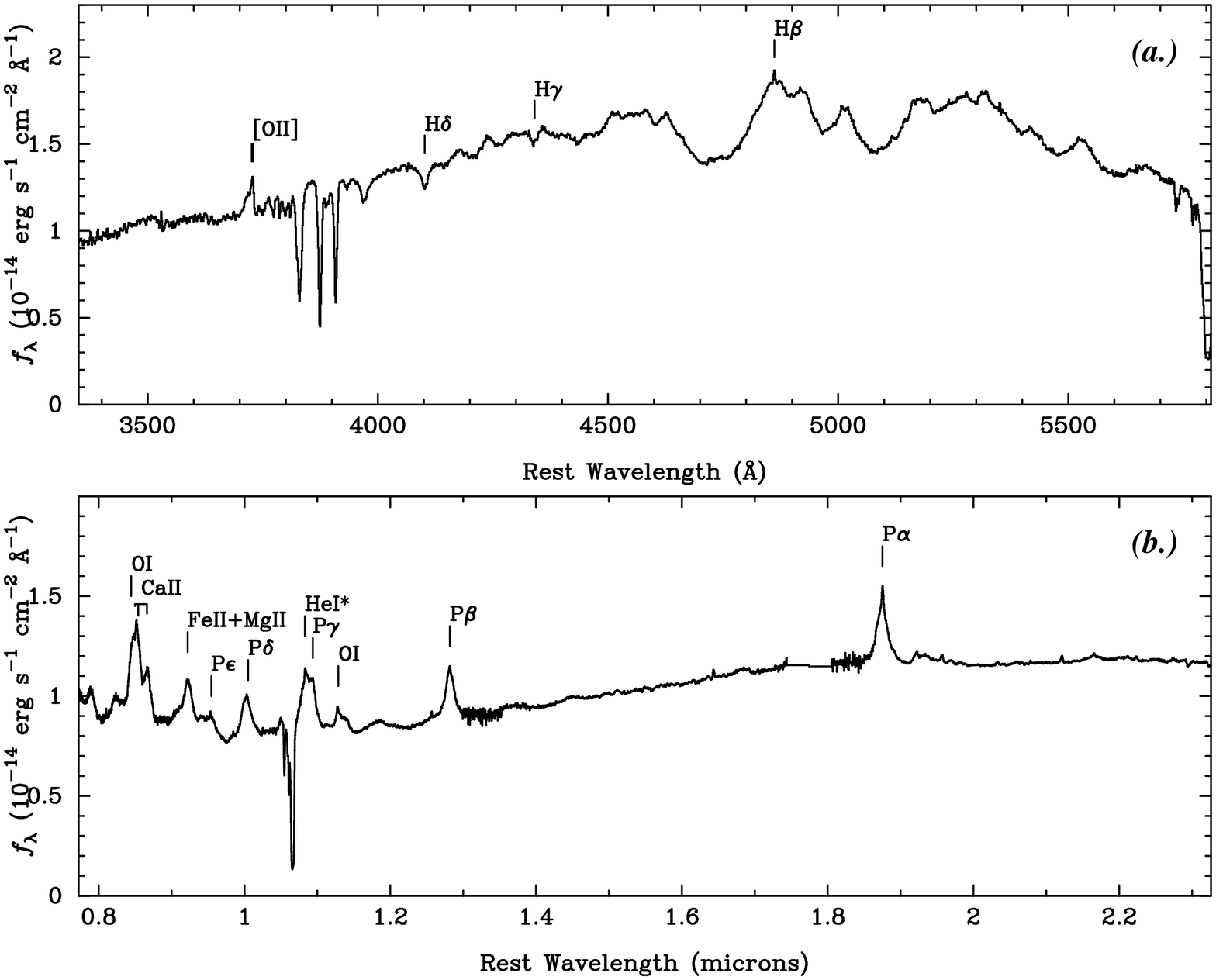}
\end{center}
\caption{ {\it (a.)} The KPNO spectrum of Mrk 231.  Principal
  emission lines  are   marked; much of the remaining line emission is
  \ion{Fe}{2} with some host galaxy emission (see also
  Fig.~\ref{fig4}).  Higher Balmer series lines appear as absorption
  lines, stemming from the contribution of the nuclear starburst (\S
  \ref{nuclear}).   {\it (b.)} The IRTF spectrum of Mrk 231. 
  Principal lines are marked; some of the remaining features originate
  in the host galaxy (see also Fig.~\ref{fig5}).    Both
  spectra have been corrected for Galactic reddening and
  redshift.  \label{fig1}}  
\end{figure}

Mrk 231 was observed using SpeX \citep{rayner03} on the NASA Infrared
Telescope Facility (IRTF) on April 24, 2010 for 30 minutes.   A
standard ABBA integration scheme was used. The A0 star HD~112623 was
used for flux and telluric corrections.  The spectra were reduced, and
the telluric correction applied, in the standard manner using Spextool
and accompanying software \citep{cushing04,vacca03}.  The resolution
measured from the arc lamp lines was about 12\AA\/ FWHM, or $\rm
330\rm \, km\, s^{-1}$ in the vicinity of the absorption line.  The
spectrum is shown in Fig.~\ref{fig1}.  The galactic reddening in the
direction of Mrk 231 is $E(B-V)=0.0095$, and we correct for that using
the CCM reddening curve \citep{ccm88}.      

We used narrow [\ion{Fe}{2}] emission lines at 12567 and 16435\AA\/ to
simultaneously check the wavelength calibration and the redshift.
These lines originate in the starburst \citep[e.g.,][]{martins13} and
therefore should provide a good estimate of the systemic velocity.
These lines yielded redshifts of 0.04248 and 0.04252, so we
corrected the spectrum using the effective redshift of
0.0425. \citet{veilleux13} mention that the redshift of Mrk~231 is
0.0422, and that value is considered to be very accurate since it is
measured from the \ion{H}{1} 21-cm absorption feature.  Our value
differs from that one by $\sim 90\,  \rm km\, s^{-1}$, and is
therefore consistent considering the resolution of the spectrum.

{ We provide the observed-frame spectrum, corrected for Galactic
reddening and normalized to $10^{-17}\rm \, erg\, cm^{-2}\,
s^{-1}\,$\AA\/$^{-1}$ in Table~\ref{irtf_spec_table}.  }

\subsection{MDM TIFKAM Observation\label{mdm}} 

Mrk~231 has been found to demonstrate variability in the \ion{Na}{1}D
line, particularly in the $\sim 8000\rm \, km\, s^{-1}$ component
\citep{boroson91,kdh92,forster95}.  To check for variability in the
\ion{He}{1}*$\lambda 10830$  absorption complex since our IRTF
observation in 2010, we obtained $J$-band spectra of Mrk 231 on the
night of UT 25 March 2013 using the TIFKAM infrared
imager/spectrograph \citep{pogge98} on the 2.4 m Hiltner telescope of
the MDM Observatory.  The spectrograph delivered a resolution of $4.24$ 
\AA\/ pixel$^{-1}$.  The slit width was $0\farcs 6$, and the effective
resolution as measured from arc lamp and night sky lines was 2.2
pixels, or $\sim 260 \rm\,  km\,  s^{-1}$ at the location of the
\ion{He}{1}* line.  We obtained 13 exposures of Mrk 231, each with an
exposure time of $300\rm \,  s$.  The A0 star HD~99966 was used for flux and
telluric corrections, and was observed $16 \times 30\rm \,  s$ at a similar
airmass to Mrk 231.  Both objects were moved along the slit in an ABBA
pattern.

The spectra were extracted in IRAF using standard procedures to trace
the spectra on pairwise-subtracted images.  We used Ar and Xe lamps to
determine the wavelength scale, and cross correlated the individual
spectra near the strong telluric absorption bands from $\lambda =
1.10$ to $1.15\ \mu$m to remove instrument flexure between exposures.
Correction for telluric absorption was done with the same method as
the IRTF spectra \citep{vacca03}.

\begin{figure}[h!]
\begin{center}
\includegraphics[width=4.0truein]{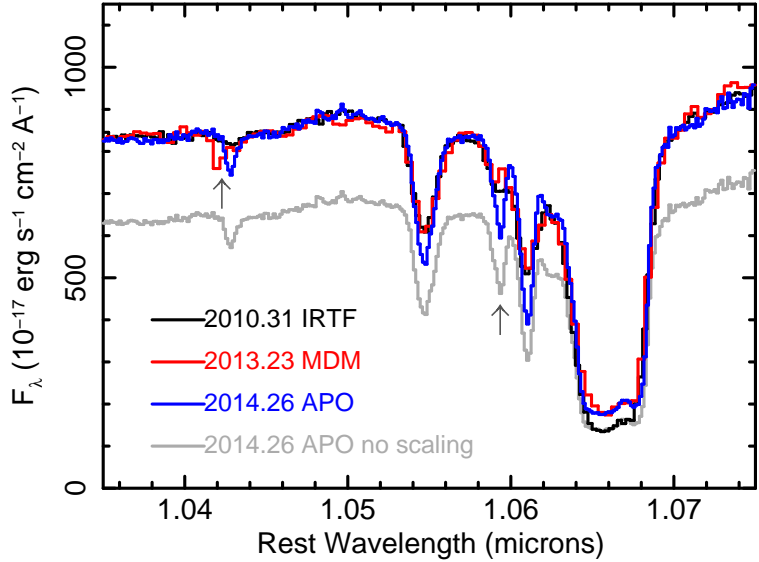}
\end{center}
\caption{ MDM TIFKAM (rescaled) and APO TripleSpec (both rescaled
  and with no scaling) spectra overlaid  on  the 2010 IRTF SpeX
  spectrum.  The  principal   \ion{He}{1}*$\lambda   10830$ trough did
  not vary between the IRTF and APO observations, while the continuum
  flux decreased by about 24\%. This is evidence that the trough is
  filled in by galaxy light.   Tentative evidence for a
  new absorption component at $v\sim   -11,520\rm   \, km\, s^{-1}$
  marked by the arrow in the MDM   spectrum is confirmed in the APO
  spectrum, with possible   deceleration to $v \sim -11,330 \rm \,
  km\, s^{-1}$.  Another component at  $v   \sim   6,620\rm \, km\,
  s^{-1}$ is observed only in the   higher-resolution APO
  spectrum, and is apparently a consequence of the higher
  resolution.  \label{fig2new}}    
\end{figure}

In Fig.~\ref{fig2new} we overlay the rescaled TIFKAM spectrum on the
IRTF spectrum.  The 
agreement between the two is excellent, with little change in the
velocities or line depths over nearly 3 years.  The depth of the
strongest feature ($v = 4,500\rm \, km ^{-1}$) is shallower in the
TIFKAM spectra; this may be a consequence of a drop in continuum flux,
as may be more prominent narrow \ion{He}{1}*$\lambda 10830$ emission
line  (\S~\ref{apo}).  

There is an indication of the appearance of an absorption feature near
$\lambda = 1.086\rm\,  \mu m$ (observed wavelength{ ; rest
  wavelength $\lambda=1.04216\rm\, \mu m$}) in the TIFKAM spectrum,
which would correspond to a velocity of $v = 11,520 \rm \, km\,
s^{-1}$ if attributed to \ion{He}{1}*$\lambda 10830$. This feature may
be marginally present in the IRTF spectrum. 
This feature cannot be a detector defect, as it is present in both the
A and B spectra.  There are no sky lines nor telluric features at this
wavelength, nor any strong starburst galaxy absorption lines
\citep[e.g.,][]{martins13}.   Therefore, we tentatively conclude that
a new, high velocity absorption feature has appeared in Mrk~231,
manifest at least in \ion{He}{1}*$\lambda 10830$.  

\subsection{APO TripleSpec Observation\label{apo}}

\begin{figure}[!h]
\epsscale{1.0}
\begin{center}
\includegraphics[width=4.0truein]{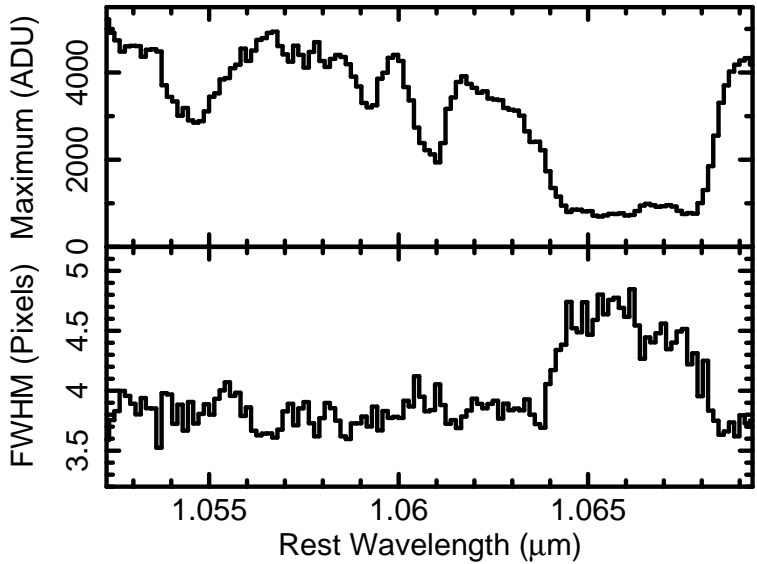}
\end{center}
\caption{ The results of measuring the profile in the spatial
  direction as a function of wavelength from the combined APO spectral
  images.   The top panel shows    the amplitude of the profile, which
  mirrors the extracted   spectrum.  The lower panel displays the FWHM
  of  the profile.  The   profile is broader in the large trough,
  suggesting that the trough   is filled in by light from the
  partially-resolved   host   galaxy.    \label{spatial}}     
\end{figure}

{ 
To confirm the presence of the new high-velocity feature, Mrk~231 was
observed on the night of 7 April 2014 using the TripleSpec infrared
spectrograph \citep{wilson04} on the 3.5-meter Astrophysical Research
Consortium telescope at the Apache Point Observatory.  A total of 16
150-second exposures were made, moving the image along the slit in an
ABBA pattern.  The $1\farcs 1$ slit was used, yielding an effective
resolution of $\sim 110\rm \, km\, s^{-1}$ (based on the widths of the
arc lamp lines)  in the vicinity of the \ion{He}{1}* line.   Ten
30-second observations of the A0 telluric standard star HD~116405 
were made at similar airmass immediately following the quasar
observation.

The spectra were extracted in a standard manner using TripleSpecTool,
a modification of SpexTool \citep{cushing04,vacca03}.  TripleSpecTool
uses the airglow emission lines for wavelength calibration.  To
account for a very small amount of flexure, wavelength calibration
solutions were computed for each pair of exposures.  Telluric
correction was performed using the same method as for the IRTF spectra
\citep{vacca03}.  

Fig.~\ref{fig2new} shows the TripleSpec spectrum overlaid on the
MDM and the IRTF spectra.  The unscaled spectrum matches the IRTF
spectrum in the bottom of the large trough, while the continuum level
is  about 24\% lower.  We interpret this as evidence that the
central engine continuum flux level of the object has decreased over
the four-year period between the observations, and that the trough is
filled in by galaxy light, principally.  The APO spectrum confirms the
presence of the new high-velocity component tentatively detected in
the MDM spectrum, although the velocity now appears to be lower, at
$11,330\rm \, km\, s^{-1}$.  

Fig.~\ref{fig2new} also shows a component at $\sim 6,620\rm \, km\,
s^{-1}$ that does not appear in either the IRTF or MDM observation.
In addition, the absorption lines appear deeper in the APO
observation.  Both of these features appear to be a consequence of the
higher resolution of the APO spectrum.  We tested this idea by
convolving the APO spectrum with a Gaussian kernel with a FWHM of
$320\rm \, km\, s^{-1}$ (three bins), approximately the value
necessary to degrade the APO $110\rm \, km\, s^{-1}$ resolution down
to the IRTF $330\rm \, km\, s^{-1}$ resolution.  When that is done,
the new feature at $\sim 6,620\rm \, km\, s^{-1}$ disappears, and the
lines become as shallow as those in the other spectra.  It is worth
noting, however, that this $\sim 6,620\rm \, km\, s^{-1}$ component
does not appear in the \citep{rupke02}  \ion{Na}{1} profile
(\S~\ref{abs_lines_2}, Fig.~\ref{fig6}) which has a higher velocity
resolution of than the APO spectrum.   

The high resolution and high signal-to-noise ratio of the APO data
yielded an additional result.  After aligning the spectral images in
the dispersion direction, we created $A-B$ pairs and aligned the
results in the spatial direction to create a single spectral trace.
We then extracted profiles in the spatial direction and determined the 
amplitude and width of the profile as a function of wavelength
(Fig.~\ref{spatial}).  The amplitude mirrors the spectrum, as it
should.  The width, however, varies.   Away from the trough, the mean
width of 3.81 pixels corresponds to a point spread function width of
about 1.5 arcseconds.  The profile within the trough is broader.  This
is most simply explained if the absorber completely covers the
continuum source, and the trough is filled in by light from the
partially-resolved host galaxy.  The broadest part of the trough has a
width of 4.66 pixels.  Assuming that the PSFs add in quadrature,
we infer that the resolved feature has an intrinsic width of 1.05
arcseconds, corresponding to 900 parsecs.  An unexpected result of
this analysis is that the longer wavelength 
portion of the trough is narrower than the shorter wavelength
portion.  In addition, the trough is not precisely flat on the bottom,
but has slightly larger amplitude at longer wavelengths.  This result
is most simply explained if a small amount of AGN or unresolved galaxy
leaks through the absorber.   Based on the difference in amplitudes,
that portion would corresponds to about $\sim 25$\% of the galaxy
light, and only $\sim 4$\% of the unobscured AGN continuum.   } 

\subsection{KPNO Observation\label{kpno}}

{ Optical spectroscopic observations} were performed during an observing run at Kitt Peak
National Observatory using the Mayall 4-meter telescope on 2011 May 6.
We used the KPC--007 grism, which has a dispersion of 1.39\AA\/ per
pixel, and resolution of 3.5\AA\/. It is also one of the bluer
gratings available, with good sensitivity up to the atmospheric
cutoff.  We observed  Mrk~231 using a blue setting with a nominal
wavelength coverage between $\sim 3200$--6000\AA\/. However, the sensitivity
dropped steeply at the blue end due to the atmospheric cutoff, and at
both ends due to vignetting.   Measurement of arc lamp lines in the 
vicinity of the features of interest indicated a resolution of $170\rm
\, km\, s^{-1}$.    

Four 15-minute observations were made for a total exposure of 60
minutes at low airmass.  Conditions were not photometric; however, the
count rates from individual spectra deviated from the mean by only
about 2.5\% at the well-exposed central part of the spectrum,
increasing to $\sim 15$--$20$\% on the vignetted edges.  We note that
the \ion{He}{1}* and \ion{Ca}{2} absorption lines were covered in the
well-exposed part of the spectrum.   Standard reduction using IRAF was
performed.   The spectrum, corrected for Galactic reddening, is in
shown Fig.~\ref{fig1}. 

The \ion{Na}{1}D absorption line fell at the very red edge
of the spectrum.  Due to vignetting and differential refraction, we
did not consider this part of the spectrum to be very reliable.
However, one feature of the data was useful.  Among the four spectra,
the red-end flux decreased steadily with time as the red image of the
object migrated out of the slit.  However, the flux in the base of the
\ion{Na}{1}D absorption line remained constant.  We take this as
an indication that the continuum filling in the base of that line
originates in the host galaxy, in particular in the nuclear starburst
known to be present \citep[e.g.,][]{davies04}.  This interpretation is
consistent with the lack of polarization of this feature
\citep[e.g.,][]{gm94,smith95, gallagher05}.  This will be discussed
further in  \S\ref{abs_lines_2}.   

We measured the redshift using a very small and narrow H$\beta$
line that lies on top of the broad H$\beta$.  This feature most likely
originates in the starburst, since narrow-line region emission in this
object is essentially undetectable.  The redshift inferred is 0.04215,
which differs from the \citet{veilleux13} value by $\sim 15\,\rm  km\,
s^{-1}$, a negligible amount considering the resolution of the
spectrum.

\subsection{Supplementary Data\label{digitize}}

A principal goal of this paper is to investigate the spectral energy
distribution from near-UV to 2.5 microns.  Our KPNO spectrum only
covers the blue optical portion, while our IRTF spectrum extends down
to 8054\AA\/ in the rest frame, and so we needed to obtain other data
from the literature to span the gap.  In addition, as discussed above,
the well-known \ion{Na}{1}D line falls at the edge of our KPNO
spectrum, and is probably not very reliable.  We therefore digitized
the optical part of the spectrum shown in Fig.\ 3 in
\citet{veilleux13}; that spectrum was obtained at Keck in 2001. That
figure did not adequately sample the \ion{Na}{1}D line, so we
digitized the absorption profile of the \ion{Na}{1}D line shown in
Fig.\ 11 of \citet{rupke02}, and applied the inferred opacity profile
to the continuum in the vicinity of the \ion{Na}{1}D.  These figures
were plotted in the observed frame, and so we shift to the rest frame
using $z=0.0422$. 

We also digitized the Mrk~231 ``5kpc'' spectrum shown in Fig.\ 3 in
\citet{rodriguez09}; this nomenclature differentiated the spectrum
from others reported in that paper that sample the outskirts of the
host galaxy. This spectrum was taken on the ISIS dual-beam
spectrograph on the 4.2m William Hershel Telescope. It has better
coverage and signal-to-noise ratio at short wavelengths than does our
KPNO spectrum, and even appears to show the \ion{Na}{1}$\lambda\lambda 
3302,3303$ absorption line corresponding to the principal BAL trough.   

In addition, we wanted to use the {\it HST} FOS spectrum to constrain
the continuum.  Since our interest was in the shape of the continuum,
rather than, e.g., detailed analysis of the absorption lines, we
simply digitized the spectrum shown in Fig.\ 1 of \citet{smith95}, and 
shifted to the rest frame using $z=0.0422$.   

\subsection{Merging the Spectra\label{merge}}

We next merged the five spectra (the KPNO spectrum, the IRTF
spectrum and the three digitized spectra) in order to obtain a broad-band
continuum.  This procedure was hampered by the fact that slit losses
make the ground-based spectra fluxes unreliable.  At the same time,
the {\it HST} observation was made in 1992, and therefore variability
could affect the relative fluxes of the spectra.  But because the {\it
  HST} FOS aperture is large (3.7''$\times$ 1.3'' effective size) and
likely did not suffer any slit losses,  we start with the {\it HST}
spectrum and proceed toward longer wavelengths, scaling segments of
spectra and joining them end to end. 

The KPNO spectrum, \citet{veilleux13} spectrum, and the
\citet{rodriguez09} spectrum overlap over part of the optical
band. Comparing these three spectra, we find that the shapes of the
\citet{veilleux13} and \citet{rodriguez09} spectra agree well
overall. However, the \citet{rodriguez09} spectrum displays an offset
of the red and blue sides at $\sim 5180$\AA\/ (rest frame) compared
with both the \citet{veilleux13} and the KPNO spectrum.  The
\citet{rodriguez09} spectrum was taken with a dual-beam spectrograph,
and we speculate that this offset originates in calibration issues at
the dichroic crossover.  Avoiding the offset region, we use the
\citet{rodriguez09} spectrum at shorter wavelengths, switching to the
\citet{veilleux13} for the necessary longer wavelength coverage.

The KPNO spectrum is slightly bluer than the \citet{rodriguez09}
spectrum ($\sim 15$\% brighter at $\sim 3600$\AA\/ when the spectra
are aligned at H$\beta$).  This is possibly due to our spectrum
including a larger fraction of the nuclear starburst, since the Balmer
absorption lines appear deeper in our spectrum.  So we do not use it
for construction of the merged spectrum, although we use it for
spectral fitting (\S\ref{nuclear}).   

Taking these considerations into account, we proceeded as follows.  The
{\it HST} spectrum overlaps with the \citet{rodriguez09} spectrum near
3150\AA\/, and we scaled and merged those two at this point.  We then
used the  \citet{rodriguez09} spectrum up to 4027\AA\/, and scaled and
merged the \citet{veilleux13} spectrum longward of that point.  The
\citet{veilleux13} spectrum extends to quite long wavelengths, with a
$\sim 2,500$\AA\/ overlap with the IRTF spectrum, but the IRTF spectrum
has better signal-to-noise ratio to short wavelengths, so we used the
scaled IRTF spectrum longward of 8223\AA\/.    

\begin{figure}[!h]
\epsscale{0.6}
\begin{center}
\includegraphics[width=4.0truein]{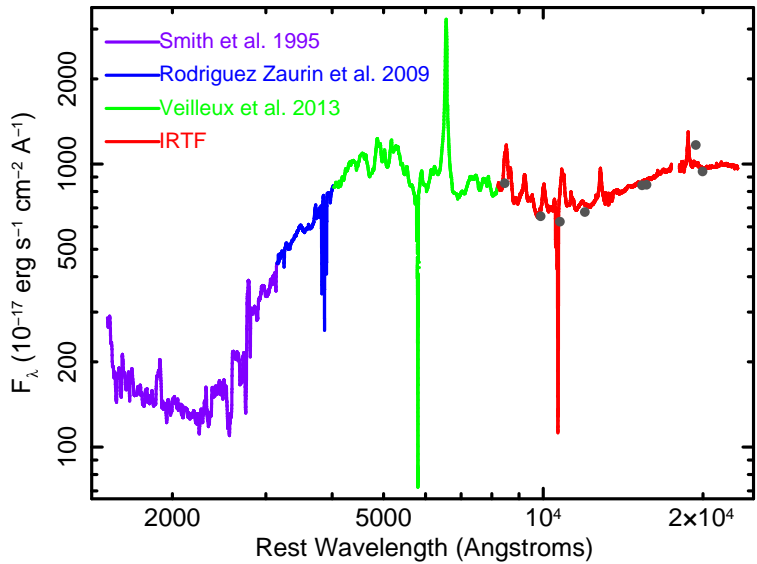}
\end{center}
\caption{\small The broad band spectrum of Mrk~231, created through merging
   digitized spectra with the IRTF spectrum.  The merging procedure
   started on the blue end with the {\it HST} spectrum, and proceeded
   segment by segment to long wavelengths (see \S\ref{merge} for
   details).  The  good agreement with the nuclear photometry points
   (gray solid  circles) suggests that the flux calibration of the
   merged spectrum is    reasonably accurate.\label{fig2}}
\end{figure}

The merged spectrum is shown in Fig.~\ref{fig2}.  Overlaid on the merged
spectra are Keck nuclear fluxes \citep[obtained using
  interferometry][]{kishimoto09} and {\it HST} \citep{kishimoto07}
fluxes obtained from NED. The long-wavelength infrared photometry
agrees well with the results of the merging process, which started
from short wavelengths. We took this convergence as evidence that our
merged spectrum is reasonably representative of the true continuum
spectrum.   

\section{Reddening\label{goobar}}

The unusual shape of the near-UV to optical continuum has been noted
previously by \citet{smith95} and \citet{veilleux13}.  As discussed by
\citet{veilleux13}, it is not consistent with any typical reddening
curve.  More specifically, the optical region of the spectrum is not
particularly red; \citet{boksenberg77} suggest $A_V=2.3$.  However,
the near UV drops dramatically, and \citet{veilleux13} show that the
HST FOS quasar composite spectrum \citep{zheng97}, normalized at $\sim
3600$\AA\/, must be reddened by a curve described by $A_V \sim 7\rm \,
mag$ to explain the steeply falling near-UV spectrum \citep[][see
their   Fig.\ 3]{veilleux13}.  { Note $A_V \sim 7$ is not, by 
itself, anomalous; for example, the radio galaxy Centaurus A is
estimated to have   $A_V\sim 14$ \citep{marconi2000}.  The property
that is both unusual  and  difficult to explain with typical
reddening curves is the apparently low reddening in the  optical
spectrum and the much higher reddening in the near UV.}   Anomalously
steep reddening has been seen in several BALQSOs \citep{hall02,
  leighly09, jiang13}, but the mechanism for this reddening remains
unexplained.   

Previous discussions of the reddening have assumed foreground
extinction in a screen with some standard ratio of total to selective
extinction, say $R_V = A_V / E(B - V) \sim 3.1$ as in
\citep[e.g.,][]{ccm88}.  Here, we offer an alternative explanation,
namely that a reddening law like that invoked to explain unusually low
values of $R_V$ around Type Ia supernovae \citep{wang05,goobar08},
provides a good description of the optical and near-UV spectrum of
Mrk~231.
 
The motivation and physics of circumstellar reddening is thoroughly
discussed in \citet{goobar08}. For a spherical scattering medium with
radius $R_{\rm CS}$ and mean free path $\lambda_{\rm
  eff}=1/n\sigma_{\rm eff}$ ($\sigma$ being a cross section), then the
attenuation is well described by the typical screen case when
$\lambda_{\rm eff}$ is much greater than $R_{\rm CS}$, similar to
extinction by dust in the interstellar medium.  But when $\lambda_{\rm
  eff} \ll R_{\rm CS}$, corresponding to multiple scatterings, more
blue photons will be removed from the line of sight, since these
photons have longer path lengths for reasonable assumptions about
wavelength-dependent scattering.  A secondary effect is that the
geometry allows photons to be added back into the line of sight.  The
radiative transfer appropriate for this case was developed by
\citet{witt92}; their Fig.\ 1 is particularly illuminating.

We use the empirical extinction parameterization derived by
\citet{goobar08}, $A_\lambda/A_V= 1-a+a(\lambda/\lambda_V)^p$
(Equation 3 in that paper), to explore whether a spherical geometry
with significant optical depth can explain the unusual reddening in
Mrk~231.  We continue to refer to this model as ``circumstellar
reddening,'' even though here we are implicitly applying it on length
scales appropriate for an active nucleus.  We note that while the
\citet{goobar08} parameterization was developed from Monte Carlo
simulations of light propagation for the $U$ (i.e., 3600\AA\/) through
$K$ photometric bands, we will extrapolate down to 2400\AA\/ in our
analysis.  Although it cannot be known whether or not the functional
form will describe this region precisely, we expect that the power law
dependence will describe the general trend, since scattering opacity
generally increases toward the blue.

\begin{figure}[!h]
\epsscale{0.6}
\begin{center}
\includegraphics[width=4.0truein]{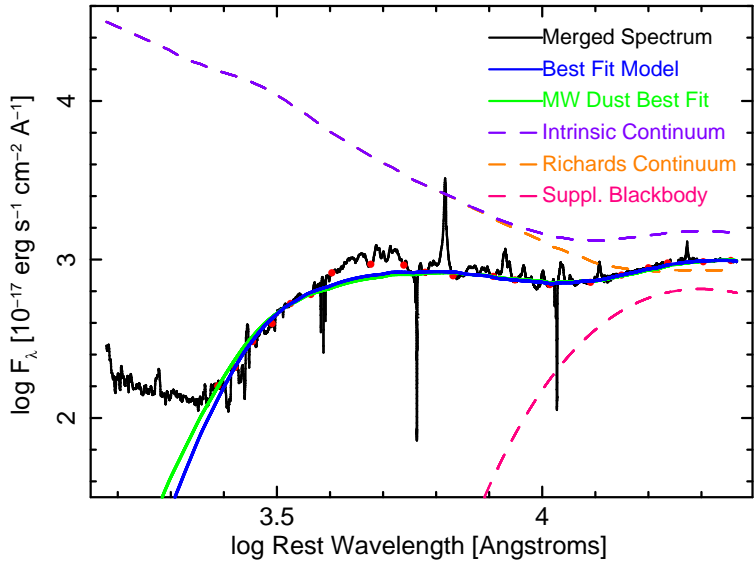}
\end{center}
\caption{\small The merged spectrum overlaid with the best fitting
   circumstellar reddening model.  The intrinsic continuum is assumed
   to be the composite quasar continuum, including the accretion disk
   and torus, derived by \citet{richards06}.  An additional black body
   with $T_{bb}=1460\rm \, K$ is necessary.  The circumstellar
   reddening is    described by $A_V$ and two shape parameters 
   \citep{goobar08}.  The best fitting 
   parameters are given in Table~\ref{table1}.  The
   best-fitting reddening-shape parameters are close to that derived
   by \citet{goobar08} for Milky Way dust; LMC-type dust does not fit  
   as well. \label{fig3}}
\end{figure}

To apply the reddening correction, we first sampled the log flux in
the merged spectrum at approximately evenly-spaced values in log
wavelength.  We avoided the strong \ion{Fe}{2} emission when possible,
and we also did not consider the continuum emission shortward of $\sim
2300$\AA\/, as it appears to be a separate component
\citep[e.g.,][]{veilleux13,smith95}.  For the intrinsic, unreddened
continuum, we used the \citet{richards06} quasar composite spectral
energy distribution, which provided good coverage over the
bandpass of interest.  This spectral energy distribution includes
nuclear continuum and torus emission, but has the host galaxy emission
removed. (Because Mrk~231 is so close, the spectroscopic slit excludes
most of the host galaxy emission except for the nuclear starburst).
We found that we needed an additional blackbody in the near infrared
(over and above the torus emission included in the \citet{richards06}
continuum) to obtain an adequate fit.  The temperature of this
component is consistent with the 3--5$\mu\rm m$ hump often seen
luminous quasars that originates in hot graphite dust at the inner
edge of the torus \citep{deo11}. The variable parameters that we
considered initially were those describing the reddening model ($A_V$
and the parameters describing the scattering and absorption properties
of the dust, $a$ and $p$; see \citet{goobar08} for details), the
blackbody temperature, and the normalizations of the blackbody and the
Richards continuum.  The figure of merit that we used to evaluate the
goodness of fit was defined as the sum of the absolute value of the
differences between the logarithmically sampled flux points and the
\citet{richards06} continuum.  We obtained two fits, one with all
parameters included in the fit, while the other with the values of $a$
and $p$ fixed at values appropriate for Milky Way dust
\citep{goobar08}.   The best fitting parameters
for these two fits are given in columns 2 and 3, respectively, in
Table~\ref{table1}, and the best fit is plotted in
Fig.~\ref{fig3}. The fit is good overall; the deviations in the
optical plausibly may be ascribed to blue continuum emission from the
nuclear starburst, and strong \ion{Fe}{2} emission. 

\begin{deluxetable}{lccr}
\tablewidth{0pt}
\tablecaption{Best Fitting \citet{goobar08} Reddening Parameters}
\tablehead{
 \colhead{Parameter\tablenotemark{a}} &
\colhead{All Parameters Free} & 
\colhead{Milky Way Dust}}
\startdata
$A_V$ & 1.54 & 1.60 \\
$a$ & 0.78 & 0.9 (f\tablenotemark{b}) \\
$p$ & $-1.72$ & $-1.5$ (f\tablenotemark{b}) \\
Inferred $R_V$\tablenotemark{c} & 2.74 & 2.79 \\
$T_{bb}$ & 1460 & 1460 \\
BB Normalization & 0.24 & 0.18 \\
Continuum Normalization & 0.85 & 0.87 \\
\enddata
\tablenotetext{a}{The first three parameters describe the extinction in the V band
and the extinction shape parameters according to $A_\lambda/A_V=
1-a+a(\lambda/\lambda_V)^p$ \citep{goobar08}.}
\tablenotetext{b}{These parameters describing the shape of the
  reddening curve were measured by \citet{goobar08}
  to be appropriate for Milky Way dust, and are fixed during the
  spectral fitting.}
\tablenotetext{c}{Computed using Eq.\ 4 from \citet{goobar08}.}
\label{table1}
\end{deluxetable}

The values of $a$ and $p$ obtained this fashion may be used to obtain
an estimate of $R_V = 2.74$, which is indicative of dust like that in
the Milky Way, in contrast with $R_V \approx 1.65$ for an LMC-like
extinction law \citep{goobar08}.  It is worth noting that SMC dust,
often used for BALQSOs \citep[e.g.,][]{gibson09}, was not
parameterized for circumstellar reddening by \citet{goobar08} because
the shape of the extinction curve was not suitable for explaining
anomalous supernovae colors.  Our best fitting parameters are close to
the Milky Way dust case; we tried the parameters provided by
\citet{goobar08} for LMC dust, and they resulted in a much poorer 
fit.

The effective $A_V$ derived from the fit is 1.5--1.6.  In
\S\ref{cloudy}, we will describe the results of {\it Cloudy}
photoionization modeling \citep{ferland13} in which we use $A_V$ as
the stopping criterion.  However, the {\it Cloudy} stopping criterion
corresponds to slab-type reddening models, while the $A_V$ derived
here is reduced due to the continuum scattered into the line of
sight.  Therefore, we needed to derive an equivalent $A_V{\rm (slab)}$ 
from our $A_V{\rm   (CS)}$.  We did that using Fig.\ 1 in
\citet{witt92}, which displays relative intensities due to direct
light, scattered light, and total light as a function of $\tau_V$ for
a dusty galaxy model.  Our best fit $A_V{\rm (CS)}=1.54$ corresponds
to an attenuation of 4.1, which corresponds to a $\tau_V{\rm
  (CS)}=1.4$.  From the figure, we found that for $\tau_V=1.4$, the
scattered light amounts to about 35\% of the total.  This implies that
for the same amount of dust and a slab geometry, the transmitted light
would have been 35\% lower, so that $A_V{\rm (CS)}=1.54$ corresponds
to $A_V{\rm (slab)}\approx 2$. 

Mrk~231 is an exceptionally highly polarized Seyfert galaxy; its
polarization properties have been discussed by \citet{smith95}, among
others. The polarization increases towards shorter wavelengths from
the optical to $\sim 3000$\AA\/, peaking at $\sim 15$\%, and dropping
toward longer wavelengths.  The high polarization underlines Mrk~231's
similarity with other BALQSOs \citep[e.g.,][]{ogle99}.  Optical
polarization increasing toward the blue, as is observed in some
Seyfert 1 galaxies, can be attributed to dust or electron scattering
combined with a reddened view of the direct continuum
\citep[e.g.,][]{wills92,leighly97}.  Our model for the reddening in
Mrk~231 assumes implicitly a spherical scattering medium.  Due to
symmetry, a perfectly spherical scattering medium will produce no net
polarization.  Therefore it may appear that our model for the
reddening is inconsistent with the relatively high polarization
observed from this object. However, we note that we would not need a
perfectly spherical scatterer and complete coverage to produce the
reddening we see; a high covering fraction would be sufficient.  Then,
intrinsic continuum may emerge through gaps in the reddening medium
and scatter towards us on e.g., dust or electrons at a larger radius
than our reddening medium.  If the gaps lie along lines to the
continuum source that are perpendicular to our line of sight, such
that the electron scattering angle approaches 90$^\circ$, high
intrinsic polarization will be produced.  Combined with the strong
reddening of the intrinsic continuum, significant net polarization
increasing toward the blue would be observed.

Our success in modeling the unusual reddening from near UV through
infrared with a modest value of $A_V{\rm (CS)} = 1.54$, corresponding
to $A_V{\rm (slab)} \approx 2$, demonstrates the plausibility of much
lower extinction values than derived by \citet{veilleux13}.
Furthermore, circumstellar reddening may be able to explain unusual
reddening curves observed in some other quasars and AGN. One object
that it may work for is IRAS~14026$+$4341 \citep{jiang13}, which shows
steep reddening like Mrk~231, and has, in addition, been reported to
show \ion{Na}{1} absorption, although it is much less strong than the
absorption in Mrk~231 \citep{boroson92}. {   \citet{jiang13},
  however, show that the unusual reddening can alternatively be 
  explained by an unusual dust-size distribution.  Specifically, large
  grains are lacking, producing less reddening in the optical part of
  the spectrum.    In either case, the interesting   question may
  be: why  does anomalous reddening occur in these two   ULIRGs and
  not in other   reddened quasars?} 

\section{Absorption Lines \label{abs_lines}} 

\subsection{Accounting for the Nuclear Starburst Emission in the KPNO
  Spectrum \label{nuclear}} 

The KPNO spectrum shows Balmer absorption lines that are a clear
indication of the presence of early-type stellar emission in the
spectrum, most likely from the nuclear starburst.  These have been
noticed in the past \citep[e.g.,][]{boksenberg77}, but are rarely
commented upon in more recent spectra.    We speculate that this
component is more prominent in our spectrum due to the use of a 
slightly large slit (1 arcsecond).  In addition, as noted above, our
KNPO spectrum appears bluer in the same bandpass than the spectrum
presented by \citet{rodriguez09}, as it should if the quasar spectrum
were contaminated by early-type stellar emission.  Our goal is to
extract the properties of the absorption lines that absorb the
central engine emission but not that of the host galaxy.  Therefore,
we need to account for the host galaxy emission before we can measure
the absorption line properties.     

To prepare the KPNO spectrum for analysis, we first renormalized it to
the merged spectrum at its long wavelength end, and then used the
best-fitting effective reddening function derived in \S\ref{goobar} to
deredden it.    We then fitted with a power law, \ion{Fe}{2}
pseudo-continuum, emission lines (principally Balmer lines),
absorption lines, and a single-burst galaxy template. We used the line
catalog from \citet{vcjv04}, convolved with a gaussian of width
$2000\rm \, km\, s^{-1}$ for the \ion{Fe}{2} pseudo-continuum.  The
Balmer lines are somewhat broad, and higher order lines have such low
contrast that they are difficult to measure independently of one
another.  Therefore, we constrained not only their width and separation,
but also their relative intensities.  We assumed that the Balmer line
emission could be approximated  by { emission from a gas in partial
local thermodynamic equilibrium \citep[PLTE;][]{popovic03}}, and
therefore their ratios are characterized by a temperature.  We found
that the temperature is not tightly constrained, and $T=5000\rm \, K$
works well.  We included Balmer lines from H$\beta$ to H9 at
3835\AA\/.  In addition, we needed an emission line that can probably 
be attributed to [\ion{O}{2}]$\lambda\lambda 3726,3729$ from the star
formation \citep[e.g.,][]{ho05}.  We assume that the absorption lines
absorb the nuclear emission (i.e, power law and emission lines), but
not the starburst continuum.   

\begin{figure}[!h]
\epsscale{0.6}
\begin{center}
\includegraphics[width=6.5in]{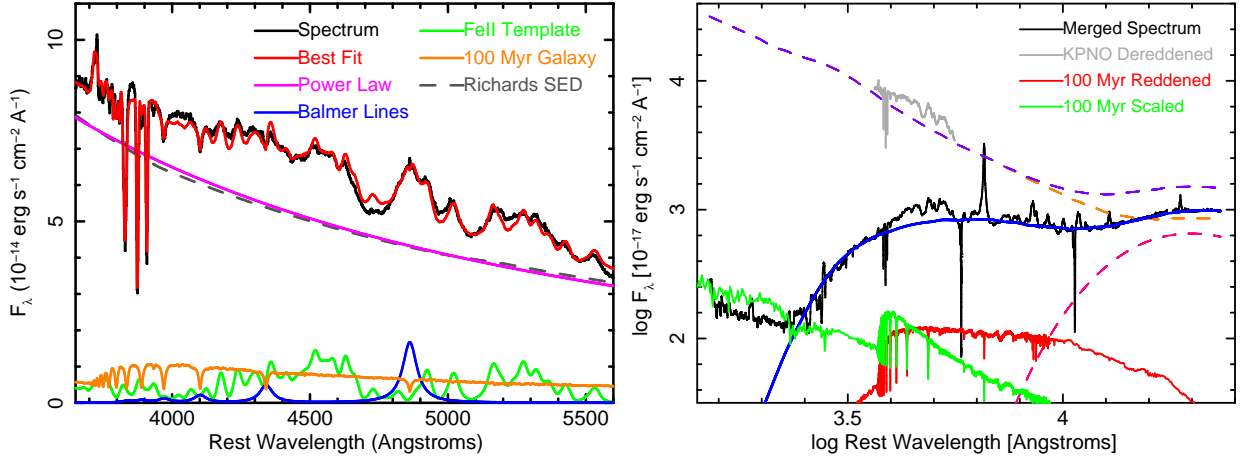}
\end{center}
\caption{\small {\it Left:} The KPNO spectrum, corrected for reddening, includes
  light from the young nuclear starburst \citep[e.g.,][]{davies04},
  modeled here using a 100~Myr single-burst galaxy spectrum.   See
  \S\ref{nuclear} for a  description of the other components. Modeling
  the galaxy spectrum allows us to determine the apparent optical
  depths of the absorption lines with respect to the nuclear
  continuum and broad-line region emission. {  {\it Right:}   The
galaxy component inferred from spectral fitting is shown in green.
An unreddened galaxy SED scaled to the amplitude of the reddened one
is shown in red.  An unreddened galaxy component this bright would
contribute unobserved stellar spectral features in the the far UV, and 
therefore we infer that the nuclear starburst is also
reddened.} \label{fig4} }  
\end{figure}

We used GALAXEV, the library of evolutionary stellar population
synthesis models computed with the isochrone synthesis code of
\citet{bc03}, for the single-burst galaxy template.  \citet{davies04}
estimated the age of the nuclear starburst to be between 10 and
100~Myr, and therefore we experimented with ready-made template
spectra for 25, 100, and 290~Myr  starbursts at $z=0.05$.  The
principal difference among these spectra distinguishable in the fitted
bandpass of the KPNO spectrum is their slope.  That is, the 25~Myr
template is the steepest, the 100~Myr is somewhat flatter, and the
290~Myr template is the flattest of the three.  In the spectral
fitting, the steepness of the galaxy spectrum is compensated by the
power law.  So, although the goodness of fit is comparable among the
three choices, the 100~Myr template yielded a power law slope and
normalization almost exactly the same as the \citet{richards06}
spectrum (see Fig.~\ref{fig4}).  This supports our contention that the
KPNO spectrum is steeper than the merged spectrum simply because of
the additional 100~Myr starburst component.   The best fitting model
is shown in Fig.~\ref{fig4}.   

The young starburst component is likely to be the same one reported
by \citet{davies04} from imaging analysis.  They inferred the presence
of a nuclear star-forming disk composed of young stars (10--100 Myr)
with length scale 0.18''--0.24'', corresponding to 150--200~pc.  It is
worth noting that the NUV spectrum was observed using FOS and 
the 4.3 aperture which subtends 3.7''$\times $3.7''\citep[effectively
  3.7''$\times $1.3'' because it overfills the
  detector;][]{smith95}. Similarly, the UV spectrum was observed using
COS with a 1.0 aperture, a circular aperture with a size of 0.86''. In
both cases, any light from the nuclear starburst would be included in
the HST spectrum.  { As noted by \citet{veilleux13}, there are no
  features originating in hot stars in the far UV spectrum,} and
therefore it appears that the nuclear starburst is also attenuated by
the circumstellar scattering.     { To investigate this point
  further, we show the inferred level of the galaxy component overlaid
  on the broad-band spectrum in   Fig.~\ref{fig4}.  The dereddened
  KPNO spectrum is shown in light grey.  Our dereddening of the KPNO
  spectrum implicitly assumes that the stellar component is reddened
  by the same amount as the AGN component.  The reddened galaxy
  component, shown in green, was obtained using the stellar component
  normalization from the spectral fitting and applying the reddening.
It is  possible, however, that only the AGN
  component is reddened, and the  galaxy component is unreddened.  We
  can approximate this situation  by normalizing the unreddened galaxy
  SED to the reddened one in the  middle of the fitting band pass,
  shown in the figure in red.  This  unreddened spectrum, extrapolated
  to the near UV, would contribute  stellar spectral features that are
  not observed.  Therefore, the  starburst is likely also reddened,
  although it may not be reddened   by the same amount as the AGN
  component.}  

The \ion{Ca}{2}$\lambda\lambda 3934, 3968$ lines and
\ion{He}{1}*$\lambda 3889$ were modeled using optical depth profiles
that have a gaussian shape.   We note that modeling the starburst
component explicitly allowed us to detect additional quasar absorption
lines attributable to higher velocity \ion{He}{1}* appearing at 3786
and 3810 Angstroms (\S\ref{abs_lines_2}, Fig.~\ref{fig4}).  These
features have been previously reported by \citet{boroson91}.  

\subsection{Extracting the \ion{He}{1}*$\lambda 10830$ Absorption
  Profile\label{hei10830}}

The IRTF spectrum is rather complicated.  There is strong emission
from Paschen lines, and the \ion{Ca}{2} IR triplet emission is very
strong and prominent.  Evidence for the nuclear starburst is also
present, including e.g., the CO 6-3 bandhead \citep{davies04} and the
narrow emission \ion{Fe}{2} lines mentioned above.   We limit detailed
analysis and discussion to the region around \ion{He}{1}*$\lambda
10830$. 

Again, we needed to estimate the starburst contribution.  This
was not easy; for example, BC03 spectra \citep{bc03} that worked so
well in the analysis of the KPNO spectrum have reduced spectral
resolution longward of $\sim 1\rm \,\mu m$. {   \citet{maraston05} 
provides model spectra with sufficient resolution; however, those
models appear to only have absorption lines, and lack the emission
lines from the starburst that are important in the
region of the spectrum around \ion{He}{1}*.}  So, to estimate the
starburst contribution, we used a near-IR template spectrum created
from the spectra of star-forming galaxies \citep{martins13} provided 
by L.\ Martins.  This template was created by averaging the near-IR
spectra of 23 low-luminosity starforming galaxies. Comparison of the 
template with the Mrk 231 spectrum revealed that various features,
such as narrow line emission and small absorption features, were
represented in both spectra.  However, as discussed by
\citet{martins13}, the spectra of the sample galaxies vary
significantly, with some objects showing brighter emission lines and
others showing deeper absorption features.   Since we don't know which
type of starburst the NIR spectrum of the Mrk~231 starburst most 
closely resembles, the results from the use of this template must be
considered representative rather than definitive.  

\begin{figure}[!h]
\epsscale{1.0}
\begin{center}
\includegraphics[width=4.0in]{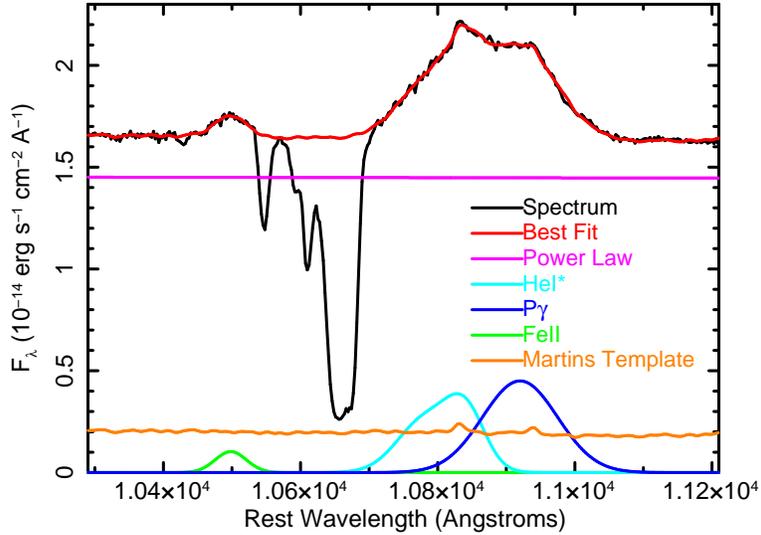}
\end{center}
\caption{\small The section of the IRTF spectrum including the
   \ion{He}{1}*$\lambda 10830$
   absorption line.  The spectrum was modeled, as discussed in \S\ref{hei10830},
   using a power-law continuum, emission lines, and a starburst
   template    \citep{martins13}.  Including the galaxy spectrum 
   explicitly in the model allows us to determine the apparent optical
   depth of the \ion{He}{1}*$\lambda 10830$ absorption line with
   respect to the nuclear continuum and broad-line region
   emission. \label{fig5}}
\end{figure}

We fitted the Mrk 231 spectrum with a model consisting of the
starburst galaxy template, a power law, and four gaussians between
10290 and 11210\AA\/, excluding regions around the absorption lines
from the fitted bandpass.  One gaussian was necessary to model
P$\gamma$, but two were required to model \ion{He}{1}*$\lambda 10830$
emission; the line apparently has a blue asymmetry.  { The bluer 
  gaussian is offset from the red one by about $1,600\rm \, km\,
  s^{-1}$, and the asymmetry may be an indication of an outflow.
  Alternatively, it could be blended emission from another,
  unidentified line, such   as has been found in PDS~456
  \citep{landt08}.}  We also included an emission line at 10498\AA\/
that may be attributed to a blend of \ion{Fe}{2}$\lambda 10491,
10502$.   The best  fit is shown in Fig.~\ref{fig5}.  It is
interesting to see how the template (which was free to vary in
normalization) nearly accounts for the continuum under the
\ion{He}{1}*$\lambda 10830$; it represents $\sim 12$\% of the total.
However, the normalization of the template in the spectral fit is most
likely being driven principally, although not wholly, by the amplitude
of the narrow \ion{He}{1}*$\lambda 10830$ and P$\gamma$ emission from
the starburst.  It is therefore not very well constrained; the error
was 9\% of the normalization for the starburst template, versus 2\%
for the power law normalization. It is worth noting that similar
relative normalizations were obtained by fitting broad-line-free
segments of the spectrum to a powerlaw plus the template.    

\subsection{Absorption Line Profiles\label{abs_lines_2}}

\begin{figure}[!h]
\epsscale{0.6}
\begin{center}
\includegraphics[width=3.5in]{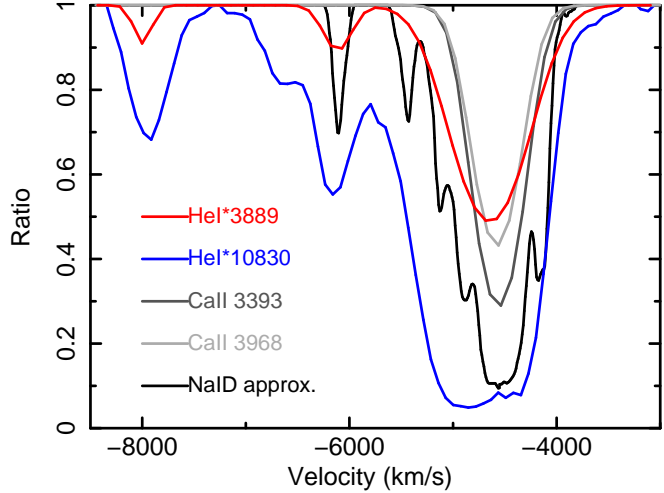}
\end{center}
\caption{\small Velocity profiles for the optical and infrared
  absorption line complexes investigated in this paper.  All lines
  were measured from our KPNO and IRTF spectra, excluding the
  starburst component (\S\ref{nuclear}, \S\ref{hei10830}), except for
  \ion{Na}{1}D, which was taken from \citet{rupke02}, and the
  approximate profile for a single component extracted as described in
  \S\ref{abs_lines_2}.  Their spectrum has higher resolution ($\sim
  65\rm \, km\, s^{-1}$) than ours (KPNO: $170\rm \, km\, s$; IRTF:
  $330\rm \, km\, s^{-1})$, and the additional structure in that
  profile is an effect of differing resolution rather than a physical
  difference.  All absorption lines have a similar velocity
  dependence; however, the \ion{He}{1}* lines, produced in the
  \ion{H}{2} part of the outflow, are characterized by higher
  velocities than the \ion{Na}{1} or \ion{Ca}{2} lines, produced in
  the partially-ionized and neutral part of the outflow. \label{fig6}}  
\end{figure}

Using the results of the spectral fitting described in \S\ref{nuclear} 
and \S\ref{hei10830}, we display the \ion{He}{1}*$\lambda 3889$,
\ion{He}{1}*$\lambda 10830$ and \ion{Ca}{2} profiles in
Fig.~\ref{fig6}. As noted above, our \ion{Na}{1}D line falls off the
edge of the detector, so here we plot the digitized profile from
\citet{rupke02}.  Their spectrum had better spectral resolution than
ours ($65\rm\, km\, s^{-1}$), revealing enhanced structure in the
profile.  Also, \ion{Na}{1}D is a doublet, with rest-frame wavelengths
of 5889.95 and 5895.92\AA\/, and effective separation of $\sim 300\rm
\, km\, s^{-1}$.  We derive an approximate velocity profile for a
single component by joining the left side of the profile, dominated by
\ion{Na}{1}$\lambda 5890$ to the right side, dominated by
\ion{Na}{1}$\lambda 5896$.  The true single component profile would be
somewhat narrower than the approximation.

The \ion{Na}{1}D profile has two principal components, including a
large one near $-4,500\rm \, km\, s^{-1}$, and a second one near
$-6,100 \rm \, km\, s^{-1}$.  A third component is sometimes seen in
this line near $-8,000\rm \, km\, s^{-1}$
\citep[e.g.,][]{boroson91,kdh92,forster95}. \citet{rupke02} find some
evidence for this component, but it is very weak.  In contrast, the
two higher-velocity components { (at $-6,100$ and $-8,000 \rm \,
  km\, s^{-1}$)}  appear stronger and more prominent in
\ion{He}{1}*$\lambda 10830$.     

Both the \ion{Na}{1} and \ion{He}{1}$\lambda 10830$ lower-velocity
components have essentially flat bottoms, a classic signature of
saturation.   The origin of the fill-in is debatable, but it quite
possibly originates in the nuclear starburst. Another piece of evidence
supporting this idea is the fact that while in other BALQSOs, where
the flux in the absorption troughs is more polarized than the
continuum \citep[as   would be the case with scattered
  light,][]{ogle99}, the polarized flux in the \ion{Na}{1} trough is
very low, essentially zero \citep[e.g.,][]{gm94, smith95,
  gallagher05}.  The level of the fill-in is slightly different in
\ion{Na}{1}D and \ion{He}{1}$\lambda 10830$; that could be real, but
could also be due to different amounts of nuclear starburst light in
the aperture.  Note that we subtracted starburst emission from the
\ion{He}{1}*$\lambda 10830$ spectrum (\S\ref{hei10830}).   

Possibly the most interesting feature is the
differences in velocities observed between the high ionization lines,
represented by \ion{He}{1}*$\lambda 3889$ and \ion{He}{1}*$\lambda
10830$, and the low ionization lines, represented by \ion{Ca}{2} and
\ion{Na}{1}.  For example, the \ion{He}{1}*$\lambda
3889$ line has a $\sim 75 \rm \, km\, s^{-1}$ higher-velocity centroid
than the \ion{Ca}{2} lines.  Furthermore, it is broader;  the
best fit width for the \ion{He}{1}*$\lambda 3889$ is $790\rm \, km\,
s^{-1}$, while the best fit width of the \ion{Ca}{2} lines is $490\rm
\, km\, s^{-1}$. Turning to the saturated lines, the FWHM points for
the principal \ion{He}{1}*$\lambda 10830$ absorption lines are $-5434$
and , $-4080\rm \, km\, s^{-1}$, while for \ion{Na}{1}, the values are
$-5144$ and $-4080\rm \, km\, s^{-1}$.  So while the red-edge
velocities line up,  the blue edges differ by $\sim 290 \rm \, km\,
s^{-1}$.  Thus, while the profiles of all the lines  discussed in this
paper are similar enough to have been produced in roughly the same
kinematic component, the \ion{He}{1}*-absorbing gas is moving faster
than the \ion{Ca}{2} and \ion{Na}{1} absorbing gas.     We will
discuss the implications of this fact on the {\it Cloudy} modeling in
\S\ref{cloudy}, and speculate on the physical conditions leading to
this result in \S\ref{picture}.

For the {\it Cloudy} modeling, we need measurements and limits of the
ionic column densities.  Note that we will limit the photoionization
analysis to the principal absorption feature centered near $-4,500\rm
\, km\, s^{-1}$.  For \ion{He}{1}*$\lambda 3889$, the \ion{Ca}{2}
lines, and \ion{He}{1}*$\lambda 10830$, we used the fitted gaussian
absorption lines to measure the apparent optical depths.  We then
integrated over the absorption profiles to obtain the ionic column
densities \citep{ss91}.  The results are given in
Table~\ref{table2}. Our measurement of log column \ion{He}{1}*$\lambda
3889$ of 15.0 is slightly smaller than the one obtained by
\citet{rupke02} of 15.3. This appears to be partially a consequence of
their curve of growth analysis method; when we integrate over a
digitized version of their profile, we obtain $15.17$.  Likewise, our
measurement of the \ion{Ca}{2}$\lambda 3934$ log column of 14.0 is
somewhat lower than theirs (14.4), possibly for the same reason.
Alternatively, absorption profile depth variability or galaxy
subtraction uncertainties may account for the differences.

\begin{table}[h!]
\centerline{Table 2. Ion Column Densities and Limits for the Principal Absorption Component}
\center{
\begin{tabular}{lcc}
\hline 
Line & log Column Density & Comment\\
\hline 
\ion{He}{1}*$\lambda 3889$ & 14.96 & measurement \\
\ion{Ca}{2}$\lambda 3934$ & 13.96 & lower limit \\
\ion{Ca}{2}$\lambda 3968$ & 14.01 & lower limit \\
\ion{Ca}{2} & 14.2 & accounting for partial covering \\
\ion{Na}{1} & 14.5 & \citet{rupke02} \\
\ion{He}{1}*$\lambda 10830$ & 14.3 & lower limit \\
\ion{H}{1} in $n=2$ & 12.3 & upper limit \\
\hline
\label{table2}
\end{tabular}}
\end{table}
\normalsize

Several of these measurements merit further comment.  { The
  \ion{Ca}{2} lines are a doublet with a ratio of the product of
  oscillator strength $f_{ik}$ and wavelength $\lambda$ of $\sim 2$.}
However,  the measured optical depth ratio is 1.48. This is a classic 
manifestation of partial covering, and therefore the column densities
measured by integrating over the apparent optical depths will yield
only lower limits on the column density.  Since we have measurements
for both lines, we use the formalism given by \citet{sabra05},
Equations 5 and 6, estimating the Ca$^+$ covering fraction and true
 column density to be 0.76 and  $2.0\times 10^{14}\rm \,
cm^{-2}$, respectively.  \citet{rupke02} estimate a similar column
density, and velocity-dependent  covering fraction between $\sim 0.7$
and 0.85.

The flat bottom of the \ion{He}{1}*$\lambda 10830$  profile suggests
that it is saturated; the estimate of the lower limit on the
\ion{He}{1}* column density obtained from this line is given in
Table~\ref{table2}.  For {\it Cloudy} modeling, we use the column
density measured from the apparently unsaturated \ion{He}{1}* at
3889\AA\/. Integrating over the line yielded a log column density of 
14.96. \citet{rudy85} estimated a similar value of $1\times 10^{15}\rm
\, cm^{-2}$.  We do not think that partial covering is significant for
this line, for the following reasons.  As discussed above, the
\ion{He}{1}*$\lambda 10830$ absorption line has a flat bottom and is
plausibly filled in by stellar continuum.  If that is true, then the 
gas containing metastable helium completely covers the photoionizing
continuum source. Since \ion{He}{1}*$\lambda 3889$ does not  appear to
be saturated, and the gas containing metastable helium completely
covers the source, and because our spectral fitting accounts for the
stellar component (\S\ref{nuclear}), then integrating over the line
yields a measurement of the true metastable helium column density
rather than a lower limit.  

As discussed above, \ion{Na}{1} falls off the end of our bandpass, and
so we use the column density estimated by \citet{rupke02} of $3.1
\times 10^{14} \rm \, cm^{-2}$.  

Finally, it is useful to estimate an upper limit on hydrogen column
density in $n=2$ from the lack of Balmer line absorption.  To do
this, we shifted the \ion{Ca}{2}$\lambda 3933$ optical depth profile to
the  H$\beta$ wavelength, and increased the optical depth until the
absorption line appeared below the spectrum minus the uncertainty.  We
then computed the corresponding Balmer column density. 
Absorption from $2s$ has a different oscillator strength
($f_{ik}=0.103$) than absorption from $2p$
($f_{ik}=0.25$). Simulations show that both states should be present
in the gas but the proportions vary.  Previously, we have tuned the
constraint to the proportions in the simulation \citep{leighly11}, but
here we  assume, for a rough estimate, that the representative
oscillator strength is the mean of the two values ($f_{ik} \approx
0.177$). This yielded  an upper limit for the hydrogen
column density in $n=2$ of $8.58\times 10^{12}\rm \,
cm^{-2}$. However, if we had had high signal-to-noise ratio spectra of
the H$\alpha$ region, we would have obtained a factor of $\sim 4.4$
times tighter constraint due to the larger values of $\lambda f_{ik}$
of H$\alpha$. Since Balmer absorption lines from the BALs have never
been reported in this well-studied object, we use an upper limit
of log hydrogen column density in $n=2$ of 12.3.  

\section{Mrk 231's Physical  Parameters\label{parameter}}

Using measurements obtained from the spectra and de-reddened SED, we
can compute various physical parameters for the quasar in  Mrk 231.
These will be needed for discussion of physical scales.  
It has been recently shown by \citet{landt13} that the Paschen lines
plus the  1-micron luminosity density can be used to obtain a
single-epoch black hole mass estimate.  The broad line FWHM is
required for this estimate. The Mrk~231 P$\alpha$ line shows an
inflection, allowing the narrow component originating in the starburst
to be modeled out. We obtained two estimates of the broad P$\alpha$
line FWHM that yielded consistent results.  The first value, $2770 \rm
\, km\, s^{-1}$, was obtained from a three-gaussian model of the
P$\alpha$ line with the narrow component removed.  The second value,
$2805 \rm \, km\, s^{-1}$, was obtained from a two-gaussian plus
starburst fit, with the starburst removed.  The 1-micron luminosity,
$\lambda L_\lambda$, was measured to be $5.5\times 10^{44}\rm \, erg\,
s^{-1}$ from the reddening-corrected intrinsic continuum shown in
Fig.~\ref{fig3}.  The black hole mass estimates from the two FWHM
values were the same to two significant figures,  so we use the value
$M_{BH} = 2.3 \times 10^8\rm \, M_\odot$.  For this value of the black
hole mass, the Eddington luminosity is $L_{Edd}=2.88\times 10^{46}\rm
\, erg\, s^{-1}$.   

We obtained the estimate of the bolometric luminosity in two ways.
First, we normalized the continuum we use for {\it Cloudy} modeling
(see \S\ref{cloudy})  to the intrinsic 2500\AA\/ flux inferred from
the dereddened intrinsic spectrum shown in Fig.~\ref{fig3}, and then
integrated over the continuum between $3.9\times 10^{11}\rm \, Hz$ to
$9.6\times 10^{4}\rm \, keV$.  The estimated bolometric luminosity was
$8.35 \times 10^{45}\rm \, erg\, s^{-1}$.  The second was obtained
using the intrinsic 5100\AA\/ flux inferred from the dereddened
intrinsic spectrum shown in Fig.~\ref{fig3}, and then applying a
bolometric correction of 10.3 \citep{richards06}.  That estimated
value of the bolometric luminosity was $7.86 \times 10^{45}\, \rm
erg\, s^{-1}$.  These
estimates are very close to one another, and we use the mean value
$8.1\times 10^{45}\rm \, erg\, s^{-1}$ henceforth.  For this value of
the bolometric luminosity, the quasar component of Mrk~231 appears to
be radiating at 28\% of the Eddington luminosity. 

\citet{tacconi02} obtain a much smaller value of the black hole mass
in Mrk~231 of $1.3 \times 10^{7}\, \rm M_\odot$, based on the stellar
dynamics, which implies a super-Eddington accretion rate.
\citet{davies04} reconsidered this result, finding that the nuclear
starburst complicates the measurement of the dynamical mass, and that
the stars have a disk-like rather than a spherical distribution,
which biases the dynamical mass estimate toward lower values.
\citet{davies04} conclude that the dynamical black hole mass is
actually $6 \times 10^8\rm \, M_\odot$,  just a factor of 2.6 times
larger than our estimate based on the Paschen line width.  

{ Finally, we needed an estimate of the size of the 1 micron
  continuum-emitting region.  The     \ion{He}{1}*$\lambda 10830$
  absorption line is deep enough that we could be certain that it
  covers both the nuclear continuum emission     and the torus
  emission.  The torus will be larger, and a conservative 
  estimate of the radius of the torus 
    contributing to the 1.083 micron continuum is $R_{\tau K}$, the
    reverberation mapping radius of K-band photometry.
  $R_{\tau K}$ is given by
\citet{kishimoto11} as $R_{\tau K}=0.47 (6\nu
L_{\nu}(5500$\AA\/$)/10^{46}\rm \, erg\, s^{-1})^{1/2} \,pc$, where
 the scaled optical luminosity $6\nu L{\nu}(5500$\AA\/) is   used
  for the UV luminosity.}  We estimated $\nu L_{\nu}(5500$\AA\/$)$
from the intrinsic continuum shown in Fig.~\ref{fig3}, and found that
$R_{\tau   K}=0.31\rm \, pc$.   

\section{Photoionization Modeling\label{cloudy}}

\subsection{General Constraints\label{general}}

Mrk~231 presents a significant challenge for photoionization
modeling, even using just the few lines accessible in the optical and
infrared spectra, as noted previously by \citet{veilleux13} and
others.  The fact that the lines have quite similar profiles, modulo
saturation,  suggests that they are all produced in gas with common
kinematics, and therefore it is reasonable to assume that it is all
part of a single physical construct.   Furthermore, while the 
saturated lines are not black, their flat bottoms and the results of
our analysis presented in \S\ref{nuclear} and \S\ref{hei10830}
suggests that the fill-in principally arises in the nuclear
starburst. Thus, the absorbing gas essentially fully covers the
continuum  emission region.  

As noted in the introduction, the most remarkable feature of the
Mrk~231 spectrum is the strong \ion{Na}{1}D line.  This line, from
neutral sodium, is common in the ISM, but rather rare in quasars. The
ionization potential of neutral sodium is only $5.14\rm \, eV$; it is
thus easily destroyed by the hard AGN continuum, and therefore it can
only  exist where it is shielded from the quasar continuum by other
gas that contains the hydrogen ionization front  \citep[e.g., see 
  also,][]{veilleux13}.    

In contrast, \ion{He}{1}* behaves as a high-ionization line.  As
discussed in \citet{leighly11} and elsewhere, \ion{He}{1}* is the
absorption from the $2s$ metastable state of neutral helium.  That
state lies $19.8\rm \, eV$ above the ground state; it cannot be
populated by collisions, but rather is populated by recombination of
He$^+$.  This means that a  photon with energy at least $24.6\rm \,
eV$ is required to form He$^+$, which in turn implies that
\ion{He}{1}* is found in the \ion{H}{2} region of quasars, along with
other high ionization lines such as \ion{C}{4}.  Furthermore, in gas
illuminated by typically hard quasar continua, the neutral helium
ionization front is coincident with hydrogen ionization front, and
therefore the \ion{He}{1}* absorption occurs in physically disjoint
gas from the \ion{Na}{1} absorption.  This fact is independent of
ionization parameter and density of the illuminated gas, and is simply
a consequence of the fact that the $24.3\rm \, eV$ photon required to
form \ion{He}{1}* would easily destroy neutral sodium.  Therefore, the
least we can say is that the gas in which the \ion{He}{1}* lines are
formed is interior (i.e., closer to the photoionizing source) to the
gas producing the \ion{Na}{1} lines.  

The second unusual and constraining property involves the fact that
the velocity for \ion{He}{1}* is slightly larger than that of
\ion{Na}{1} and \ion{Ca}{2}.  This fact by itself is not unusual;
high-ionization lines often have higher velocities than low-ionization
lines \citep[e.g.,][]{voit93}.  However, in those cases, there are
also usually differences in profiles and evidence for partial
covering, so that, e.g., the low ionization lines could be produced in
dense cores with relatively low velocity, while the high ionization
lines are produced in a thinner wind possibly ablated from the cores and
accelerated.  For Mrk~231, the similarity of the profiles and the
evidence for full covering, specifically for the \ion{He}{1}*$\lambda
10830$ and \ion{Na}{1}D lines, implies that the same gas is
responsible for both sets of lines.  Other full-covering situations
that might be similar would be, e.g., stellar winds and
supernovae. However, in those cases, the flow is homologous, that is,
$v \propto r$.  This is opposite of  the situation for Mrk 231, where 
the gas at a smaller radius (absorbing in \ion{He}{1}*) is moving
faster than the gas at a larger radius (absorbing in \ion{Na}{1} and
\ion{Ca}{2}).  This physical situation suggests perhaps  that a shock
would be produced at the interface.

All of these facts imply that the absorption lines in Mrk 231 are
sufficiently different from other BALQSOs that we cannot take the
standard analysis approach, i.e., running a grid of models, varying
density, ionization parameter, and column density, as well as
metallicity and spectral energy distribution, and then determining the
location in parameter space that best fits the absorption lines.  Our
approach is instead suggested by the physical picture laid out in the
next section.  

\subsection{A Physical Picture and Cloudy Simulation Setup\label{picture}}

Mrk 231 has a well-known nuclear starburst.  Detailed analysis by
\citet{davies04} reveal that the nuclear starburst has a disky
configuration with length scale of 150--200 pc and an age of
10--100~Myr.  Our analysis in \S\ref{goobar} shows that the quasar
continuum suffers extinction in a shell of dust that subtends a
large solid angle. We suggest that the nuclear starburst produces the
dust performing the scattering, and that the nuclear starburst may
also be the origin of the gas in which the \ion{Na}{1}D and
\ion{Ca}{2} absorption is occurring.  However, a starburst would
produce, by itself, very modest outflow velocities \citep[hundreds of
  $\rm km\,   s^{-1}$;   e.g.,][]{rupke05}.  So we suggest that the 
acceleration originates from a a continuous quasar outflow impacting
upon and accelerating, over time, the gas being processed by the
starburst.  The quasar outflow is the origin of the \ion{He}{1}*
absorption, while the accelerated processed gas is the origin of the
\ion{Na}{1} and \ion{Ca}{2} absorption.  This scenario can explain why
the \ion{He}{1}* lines have higher velocity than the low-ionization
lines. 

We note that quasar feedback, thought to be necessary to shut down
star formation during the co-evolution of black holes and quasars,
posits an interaction between quasar outflows and surrounding
star-forming gas.  It therefore may be that we are observing an
example of feedback in action.  

This scenario also motivates what we have found by experimentation to
be a key requirement in the photoionization model: a density increase
between the \ion{H}{2} region, where the \ion{He}{1}* absorption
occurs, and the partially-ionized zone, where the \ion{Ca}{2} and
\ion{Na}{1}D lines are formed.  This is because not only does Mrk~231
show strong \ion{He}{1}* absorption, but the inferred column density
of this line is quite large.  As discussed in \citet{leighly11}, 
a high column of metastable helium requires a high column of He$^+$.
A large column density of He$^+$ is attained by a relatively high
ionization parameter.  In contrast, neutral sodium favors a low
ionization parameter.   Those two conditions can be met in the same
slab of gas if there is an increase in density between \ion{H}{2}
region and the partially-ionized zone. Further, we assert that
without the density gradient, Mrk~231 perhaps would resemble other
FeLoBALs, similar to e.g., FBQS~1151$+$3822 \citep{lucy14}, with
\ion{He}{1}*, magnesium and  \ion{Fe}{2} lines, but no very low
ionization lines like \ion{Ca}{2} or \ion{Na}{1}.   
 
Physically, the density gradient or step could result from the BAL
outflow impacting the starburst-processed gas, scooping it up and
accelerating it. Perhaps something like the cloud-crushing scenario
proposed by \citet{fg12}, in which shocked gas cools, could be
operating, with the difference that a large covering fraction should
be produced.   In principle, a dynamical model for the envisioned
scenario could be constructed; however, that is beyond the scope
of this paper, and it is not clear that such a model could be
constrained usefully with the small number of lines measured.

Instead, we present a simple (i.e., toy) model to see if we can produce
the column densities and limits that we measure, and explore the
conditions required.  We start with an assumption of constant pressure
to produce the density gradient.  We also assume solar abundances.
While Mrk~231 appears to have an X-ray weak spectral energy distribution
{ \citep{teng14}}, we start with a typical quasar SED in order to
examine the influence of this parameter, and investigate an X-ray
weak SED in \S\ref{continuum}.  If the absorption is occurring in the
starburst, dust is probably present in the gas.  The analysis
presented in \S\ref{goobar} shows that, at least in the line of
sight, the extinction is relatively modest, with $A_V(slab)\sim
2$. Therefore, we use, in our simulations,  $A_V=2$ as a stopping
criterion.  Two other parameters describing the gas are important: the
dust-to-gas ratio and the depletion of various elements from the gas
phase into dust.  

\S\ref{sims} describes the {\it Cloudy} simulations and results for this
simple model. The results are suggestive, but some aspects would be
difficult to explain physically.  \S\ref{beyond} then describes the
results of modifying the simple model assumptions one by one.  Finally,
\S\ref{optimize} describes the results for a optimizing all of the
parameters described in \S\ref{beyond}.

\subsection{Simulations and Analysis\label{sims}}

We used {\it Cloudy 13.02} \citep{ferland98, ferland13}.  For the
initial spectral energy distribution, we used the so-called {\it
  Cloudy} AGN continuum with the same parameters as used 
by \citet{korista97}\footnote{The {\it Cloudy} command for this
  continuum is ``AGN Kirk''.  The model consists of a powerlaw with a
  slope of $\alpha_{uv}=-0.5$, an high energy exponential cutoff with
  $\log T=6$, an X-ray power law with a slope $\alpha_{x}=-1.0$, and
  $\alpha_{ox}=-1.4$}.  Default {\it Cloudy} solar photospheric 
abundances, taken from various sources described in the {\it Hazy}
manual included with the {\it Cloudy} distribution, were used.  We then
modified those abundances using the approximate depletion scheme
described below. Given that the reddening inferred in \S\ref{goobar}
was best described by Milky Way dust, we chose the {\it Cloudy} built-in ISM
dust \citep{dl07}, and as noted above, used $A_V=2$ as our stopping
criterion.  That is, the gas slab thickness was truncated by the
software when $A_V=2$.   

The Milky Way galaxy is characterized approximately by a constant
dust-to-gas ratio.  But in the chaotic environment of a quasar, a
Milky Way dust-to-gas ratio can not be guaranteed.  Therefore,
dust-to-gas ratio was included as a parameter, where, for example, a
parameter value of 0.1 means that the dust-to-gas ratio was 10\% of
that in the ISM.  

If dust is present, then some elements should be depleted onto the
dust grains. It may be expected that the dust-to-gas ratio and the
depletion should be coupled parameters; the larger the dust-to-gas
ratio, the more metals should be removed from the gas phase.  But they
can be approximately independent if the dust is clumpy.  In that case,
significant metals can be removed from the gas by depletion onto dust,
but if those dust clumps are larger than the relevant wavelengths,
they may provide little reddening.  We started with {\it Cloudy}'s
built in depletion scheme\footnote{The   full depletion   scheme is
  given on page 77 of   the   {\it Cloudy} 13.02 manual {\it     Hazy}
  and references therein.}; for example, it predicts, for a standard
dust-to-gas ratio, that calcium should be depleted by a factor of
$10^{-4}$, and sodium should be depleted by a factor of 0.2.   We then
defined a depletion parameter that ranged between zero and 1, and
gives the logarithmic fraction of the depletion.  Specifically, the 
standard calcium depletion by a factor of $10^{-4}$ would be assigned
an abundance parameter equal to 1, while an abundance parameter of
$0.5$ would correspond to a depletion by a factor of $10^{-2}$.  This
is simply a parameterization and there is no physical basis behind it.
But given the uncertainty of dust depletion factors, especially
potentially occurring in the AGN/starburst environment of Mrk~231, it
served our purpose.  

We explored ionization parameters between $\log U=-1.5$ and $\log
U=0.5$.  We used a constant pressure to produce the density gradient,
and specified the hydrogen density $\log n$ at the illuminated side of
the slab.  We explored values between $\log n=3$ and $\log n=5.5$ $\rm
\, [cm^{-3}]$.  Thus our simulation grid was described by four
parameters: the ionization parameter $\log U$, the density $\log n$,
the dust-to-gas ratio parameter, and the depletion parameter.   

The simulations produced predicted column densities of the key ions:
neutral sodium, metastable neutral helium, Ca$^+$, and hydrogen in
$n=2$.   The simulation column densities were compared with the
measured values given in Table~\ref{table2} using a modified figure of
merit ($MFOM$).  Previously, we have used a figure of merit defined as
the sum of the absolute value of the difference between the measured
log column density and the simulation results weighted by the
measurement uncertainties \citep[e.g.,][]{leighly04, casebeer06,
  lucy14}.  We have found that this figure of  merit performs better
than $\chi^2$ in the face of outliers, although it is not easy to
interpret statistically. Here,  however, we needed to combine limits
with measurements with error bars.  Therefore, we used a modified
figure of merit (MFOM) defined as before, except that we assigned a
contribution to the sum equal to zero for an ion if either falls
within the range defined by the measurement uncertainties, or is
consistent with the upper or lower limit as appropriate for that ion.
Therefore, any simulation which yielded an $MFOM$ equal to zero has
all four column densities in agreement with the measurements and the
limits.     

We used a lower limit on neutral sodium of $14.5$, a lower limit on
\ion{He}{1}* of $14.8$, an upper limit on hydrogen in $n=2$ of $12.5$,
and bounds on Ca$^+$ between 14.1 and 14.3.  The sodium and hydrogen
limits are the same as in Table~\ref{table2}, but our choices for
\ion{He}{1}* and Ca$^+$ require some explanation.  First, we consider
\ion{He}{1}* a lower limit, given that we were not able to solve for
partial covering for this line, although based on the  discussion
above, we think that the covering fraction is nearly 100\%.  Our best
measured value for \ion{He}{1}* is 14.96, and, in principle, that
should be the lower limit.  However, we relaxed that the limit to
$\log N_{HeI*} > 14.8$, because of a limitation in the {\it Cloudy}
modeling software.  Physically, the \ion{H}{2} region of  the absorbing
gas may be dust free if it is a wind that originates from the quasar
(see \S\ref{picture}). However,  {\it Cloudy} cannot accommodate a
dust-to-gas ratio that varies through the slab.   As will be discussed
below,  the dust unavoidably included in the simulation in the
\ion{H}{2} region suppresses production of metastable helium, and
therefore, the \ion{He}{1}* column produced by the simulation would be
higher if dust were not present in the \ion{H}{2} region.
Furthermore, based on previous studies of \ion{He}{1}* in the FeLoBAL
quasar FBQS~J1151$+$3822 \citep{leighly11, lucy14}, we know that we
can attain column densities of metastable helium in the
\ion{H}{2} region for these values of ionization parameter, if dust
were not present. Second, although we have a good measurement of the
Ca$^+$ column of $14.2$ including partial covering analysis
(\S\ref{abs_lines}), the measurement is uncertain due to possible
systematic errors and model dependencies; for example, the depth of
the \ion{Ca}{2} lines in the KPNO spectrum depends on the fractional
contribution of the starburst, which, in our spectral fitting,
depended on the age of the assumed starburst spectrum, and that was
degenerate with the power law slope. Therefore, we assigned $MFOM=0$
when the Ca$^+$ value lay between 14.1 and 14.3.  

\begin{figure}[!h]
\epsscale{1.0}
\includegraphics[width=6.5in]{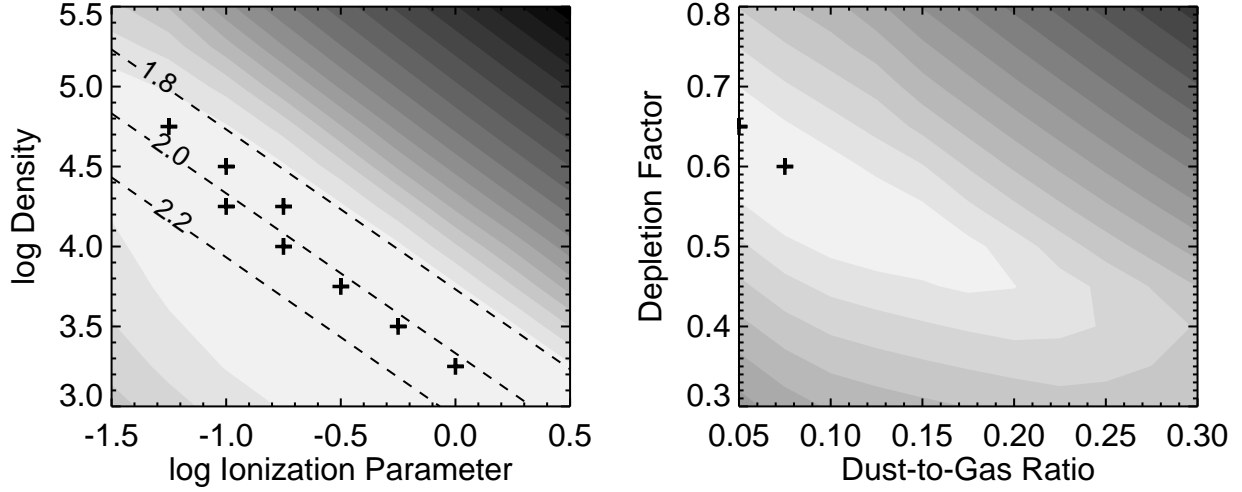}
 \caption{\small Contours of the modified figure of merit (MFOM) as defined
   in \S\ref{sims} as a 
   function of simulation parameters, for a fixed $A_V=2$, in
   orthogonal directions to the characteristic  solution
   ($\log(U)=-1.5$, $\log(n)=3.75$, dust-to-gas scale factor equal to
   0.075, and depletion scale factor of 0.6).  The plus
   signs show where $MFOM=0$, i.e., where the solutions are consistent
   with the limits and bounds (\S\ref{sims}).  The dashed lines show
   contours of constant log distance of the absorber from the central
   engine, in  units of parsecs.  So, the solutions imply that the
   absorber lies  $\sim 100\rm  \, pc$ from the central engine, near
   the location of    the circumnuclear starburst
   \citep[e.g.,][]{davies04}.  \label{fig7}} 
\end{figure}

Since there are multiple points in parameter space where $MFOM =0$, we
did not assign a best fit based on the value of $MFOM$.  Instead, we
identified a characteristic solution defined  by the $MFOM=0$
grid point that lay at the minimum of the sum of the differences
between each $MFOM=0$ solution and every other one, normalized by the
grid  spacing.   For this toy model, the characteristic solution turns
out to be $\log U=-0.5$, $\log n=3.75$, dust-to-gas ratio parameter
equal to 0.075, and depletion parameter equal to 0.6.  Fig.~\ref{fig7}
shows contours of $MFOM$ in terms of  our four parameters, orthogonal
to this point.  The parameter values where $MFOM=0$ are
marked. Table~\ref{table3} gives the ranges and the values for the
characteristic solution, which will be discussed in detail in
\S\ref{char_sol}. 

\begin{deluxetable}{lcccccc}
\footnotesize
\rotate
\tablewidth{0pt}
\tablecaption{{\it Cloudy} Modeling Results}
\tablehead{
\colhead{} & \multicolumn{2}{c}{Constant Pressure} &
  \multicolumn{2}{c}{Density Jump = 0.6\tablenotemark{a}} & \multicolumn{2}{c}{Density
    Jump = 1.4\tablenotemark{a}} \\
\colhead{Parameter} & 
\colhead{Range\tablenotemark{b}} & 
\colhead{Characteristic} &
\colhead{Range\tablenotemark{b}} & 
\colhead{Characteristic} & 
\colhead{Range\tablenotemark{b}} & 
\colhead{Characteristic}\\
\colhead{} & 
\colhead{} & 
\colhead{Solution\tablenotemark{c}} &
\colhead{} & 
\colhead{Solution\tablenotemark{c}} & 
\colhead{} & 
\colhead{Solution\tablenotemark{c}}}
\startdata
\multicolumn{7}{c}{Simulation Parameters} \\
$\log U$\tablenotemark{d} & $-1.25$ -- 0.5 & $-0.5$ & $-1.25$ -- 0.75 & $-0.25$  & $-1.5$ -- 0.75 & $-0.5$ \\
$\log n$ [$\rm cm^{-3}$]\tablenotemark{e} & 3.0 -- 4.75 & 3.75 & 3.0 -- 5.5 & 4.25 & 3.0 -- 5.5 & 3.75 \\
Dust-to-Gas Ratio Parameter\tablenotemark{f} & 0.05 -- 0.1 & 0.075 & 0.05 -- 0.175 & 0.075 & 0.05 -- 0.3 & 0.125 \\
Abundance Parameter\tablenotemark{g} & 0.5 -- 0.65 & 0.6 & 0.45 -- 0.55 & 0.5 & 0.4 -- 0.6 & 0.5 \\ 
\hline
\multicolumn{7}{c}{Column Densities [$\rm cm^{-2}$]} \\
log Total Hydrogen Column & 22.6 -- 22.9 & 22.7 & 22.3 -- 22.9 & 22.7 & 22.1 -- 22.9 & 22.5 \\
log \ion{H}{2} Region Column  & 21.6 -- 22.2 & 21.9 & 21.1 -- 21.8 &  21.3 & 21.1 -- 22.0 & 21.7 \\
log \ion{H}{1} Region Column  & 22.4 -- 22.8 & 22.6 & 22.3 -- 22.9 & 22.7 & 22.0 -- 22.9 & 22.4  \\
\ion{Na}{1} & 14.5 -- 15.1 & 14.85 & 14.5 -- 14.9 & 14.6  &  14.5 -- 15.2 & 14.6 \\
\ion{He}{1}* & 14.8 -- 14.9 & 14.83 & 14.8 -- 15.3 & 15.0 & 14.8 -- 15.4 & 15.2 \\
\ion{Ca}{2} & 14.1 -- 14.3 & 14.14 & 14.1 -- 14.3 & 14.3 & 14.1 -- 14.3 & 14.2 \\
\ion{H}{1} n=2 & 11.8 -- 12.3 & 11.95 & 10.6 -- 12.3 & 11.3 & 10.4 -- 12.3 & 11.3 \\
\ion{Mg}{1} & 15.0 -- 15.4 & 15.2 & 14.7 -- 15.0 & 14.8 & 14.6 -- 15.2 & 14.8 \\
\ion{Mg}{2} & 17.6 --17.9 & 17.7 & 17.6 -- 18.1 & 17.9 & 17.2 -- 18.0 & 17.6 \\
\ion{Fe}{2} Low\tablenotemark{h} & 16.7 - 17.0 & 16.8 & 16.9 -- 17.3 & 17.2  & 16.6 -- 17.3 & 16.9 \\
\ion{Fe}{2} High\tablenotemark{i} & 13.2 -- 14.5 & 13.8 & 11.6 -- 14.1 & 12.8 & 11.5 -- 14.6 & 12.7  \\
\ion{Mn}{2} & 15.2 -- 15.5 & 15.3 & 15.2 -- 15.7 & 15.5 & 14.9 -- 15.6 & 15.2  \\
\ion{C}{4} & 17.2 -- 17.4 & 17.3 & 16.1 -- 17.1 & 16.4 & 16.1 -- 17.3 & 17.0  \\
\ion{N}{5} & 16.8 -- 17.0 & 17.0 & 15.5 -- 17.3 & 16.0 & 15.5 -- 17.5 & 17.1 \\
\hline 
\multicolumn{7}{c}{Kinematic and Other Parameters} \\
$\lbrack$\ion{O}{3}$\rbrack$ Equivalent Width (\AA\/)\tablenotemark{j}
& 150 -- 270  & 190 & 150 -- 200  & 190 & 120 -- 200 & 150 \\
Outflow Radius (pc) & 62 -- 150 & 110 & 13 -- 230 & 54 & 13 -- 230 &
100  \\
Outflow Mass Flux ($\rm M_\odot \, yr^{-1}$) & 600 -- 2830 & 1410 &
170 -- 4400 & 700 & 55 -- 4400 & 740  \\
log Kinetic Luminosity [$\rm \, erg\, s^{-1}$] & 45.6 -- 46.3
& 46.0 & 45.0 -- 46.4 & 45.6 & 44.5 -- 46.5 & 45.7 \\
Kinetic / Bolometric Luminosity & 0.47 -- 2.2 & 1.1 & 0.13 -- 3.5 &
0.5 & 0.04 -- 3.5 & 0.6 \\
\hline
\enddata
\tablenotetext{a}{The density jump is defined as the logarithmic
  factor by which the density increases at the hydrogen ionization
  front (\S\ref{const_press}).}
\tablenotetext{b}{The range of the parameter for solutions where $MFOM=0$.}
\tablenotetext{c}{The characteristic or typical value (see
  \S\ref{sims} for a description of how this parameter is calculated).}  
\tablenotetext{d}{Simulation range: $-1.5 < \log U < 0.5$ for constant
pressure, $-2.0 < \log U < 1.0$ for density step simulations.}
\tablenotetext{e}{Simulation range: $3.0 < \log n < 5.5$.}
\tablenotetext{f}{The dust-to-gas ratio parameter is the fraction of
  the normal dust to gas ratio for the ISM.  Simulation range: $0.05$
  to $0.3$. See \S\ref{sims} for details.}
\tablenotetext{g}{The abundance parameter is the logarithmic fraction of the normal
ISM depletion.  Simulation range: $0.3$ to $0.8$. See \S\ref{sims} for
details.}
\tablenotetext{h}{Column density of Fe$^+$ atoms populating energy levels up to
0.12~eV above the ground state. A strong feature in the near UV from
transitions from these levels appears between 2585 and 2631\AA\/.  See
\citet{lucy14} for details.}
\tablenotetext{i}{Column density of $Fe^+$ atoms populating energy levels between
0.98 and 1.1 eV above the ground state.  A strong feature in the near
UV from these levels appears between 2692 and 2773\AA\/.  See
\citet{lucy14} for details.}
\tablenotetext{j}{Assumes global covering fraction of 0.2.}
\label{table3}
\end{deluxetable}
\normalsize

This figure has several interesting features.  First, our
constant stopping criterion of $A_V$ plus dust-to-gas ratio
specifies a hydrogen equivalent column density.  Thus, the $\log U$ vs
$\log n$ graph (Fig.~\ref{fig7}, left side) shows values for a
constant $\log N_H=22.7 \rm \, cm^{-2}$. Second, since $U=\phi/nc$,
where $\phi$ is the photoionizing flux, contours of constant
photoionizing flux run diagonally across the plot, along the contours.
Constant photoionizing flux occurs at constant radius. Thus, our
$MFOM=0$ solutions all lie approximately the same distance from the
central engine.   

\begin{figure}[!h]
\epsscale{1.0}
\begin{center}
\includegraphics[width=3.9in]{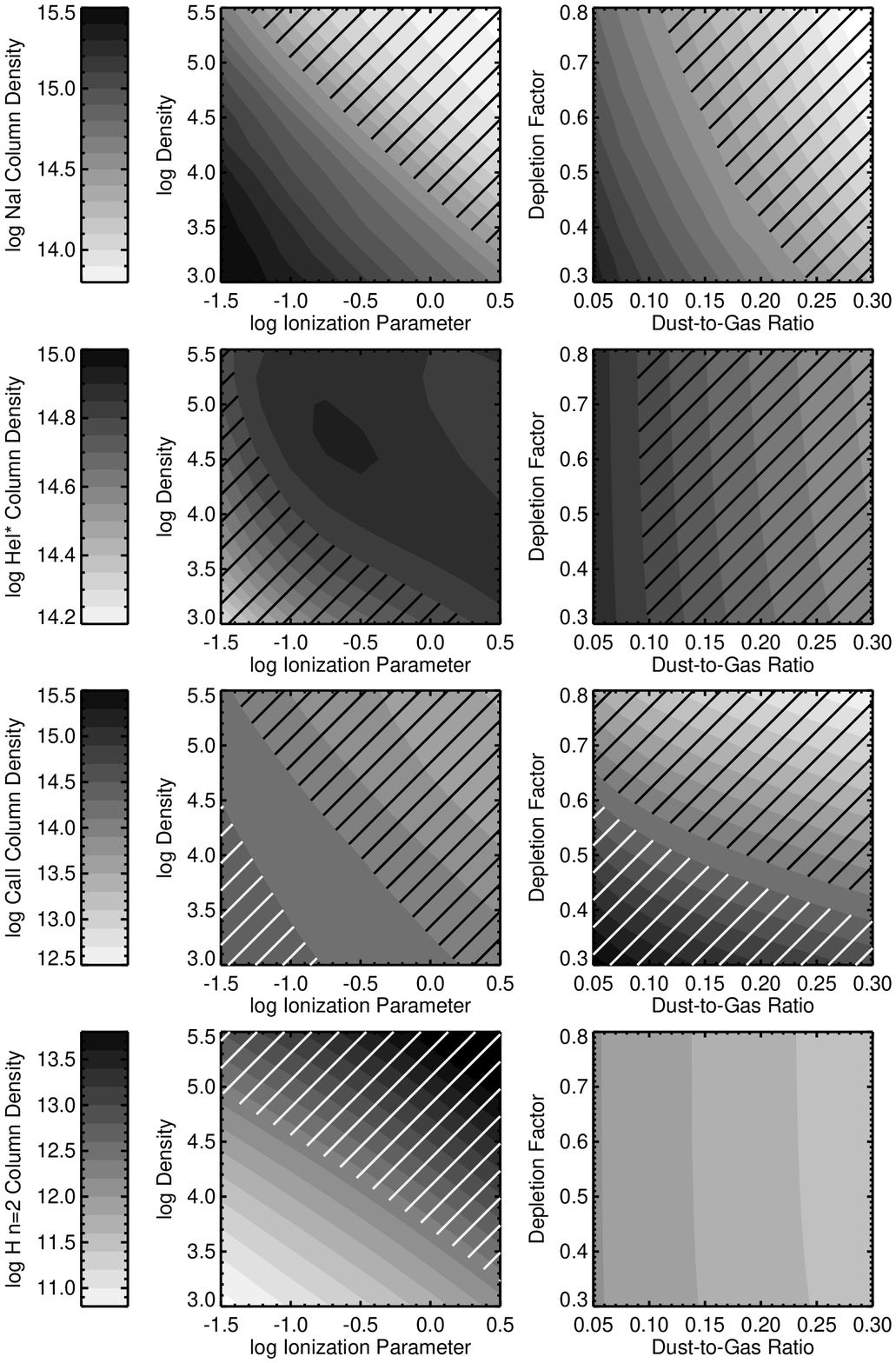}
\end{center}
 \caption{\small Contours of ionic column density as a 
   function of the simulation parameters, for a fixed value of
   $A_V=2$, in orthogonal directions from the characteristic solution
   ($\log(U)=-1.5$, $\log(n)=3.75$, dust-to-gas scale factor equal to
   0.075, and depletion scale factor of 0.6).  From the top to the
   bottom, we should results for \ion{Na}{1}, \ion{He}{1}*,
   \ion{Ca}{2}, and \ion{H}{1} in $n=2$.  The hatched areas show
   regions not consistent with the limits or bounds, where black
   hatches indicate a lower limit, and white hatches indicate an
   upper limit.  See  \S\ref{sims} for discussion.  \label{fig8}} 
\end{figure}

Fig.~\ref{fig8} shows contours of ionic column density as a function
of the simulation parameters, also orthogonal to the characteristic 
solution.  These graphs show how the various lines constrain the
simulation parameters.  Neutral sodium is strongly dependent on the
ionizing flux.  For a fixed column density, the lower the
photoionizing flux, the more likely that a predominately neutral zone,
necessary for neutral sodium to survive, will be present at the back
of the slab. Neutral sodium depends inversely on the dust-to-gas
ratio; the larger the gas fraction, the thicker the gas for the fixed
value of $A_V$, and thus the more likely a neutral zone will be
present at the back of the slab.  

\ion{He}{1}* shows interesting ionization parameter/density
dependence.  For low values of photoionizing flux (low values of $\log
U$ and $\log n$), the \ion{He}{1}* first increases as photoionizing
flux increases.  As discussed in \citet{leighly11} and elsewhere, the
amount of \ion{He}{1}* depends on the thickness of the He$^+$ region,
which increases as $\phi$ increases.   But for higher values of
$\phi$ (high values of $\log U$ and $\log n$), \ion{He}{1}*
decreases, the opposite of what would be expected for dust-free gas  
\citep[e.g.,][]{leighly11}.  This happens because, as $\phi$
increases, the He$^+$ region is shifted toward the back of the
slab, where the accumulated extinction, proportional to depth in the
slab, is higher.  The increased extinction removes photons 
in the helium-ionizing continuum, and so there is less He$^+$
produced, and consequently less \ion{He}{1}*.  There is no dependence
on depletion, since helium is not depleted onto grains, and the
dependence on dust-to-gas ratio mirrors the column density variation
with this parameter, coupled with the extinction.  

The low-ionization ion Ca$^+$ has similar dependence on ionization
parameter and density as neutral sodium.  It has, however, stronger
dependence on depletion than any of the other ions.  This is because
calcium is the most strongly depleted onto dust of all elements. 

Finally, hydrogen in $n=2$ strongly depends on photoionizing flux.
The lack of Balmer absorption lines is generally a good density
indicator; we used this to obtain a density upper limit for
FBQS~J1151$+$3822 in \citet{leighly11} and \citet{lucy14}. In this
case, however, the combination of \ion{Na}{1} and \ion{He}{1}* provide
stronger constraints on the photon flux and therefore density.

Thus, the lines play complementary roles in constraining parameter
space.  The neutral sodium, Balmer lines and \ion{Ca}{2} push the
solution toward lower photon fluxes, and therefore lower values of
ionization parameter and density.  In contrast, the \ion{He}{1}*
pushes the solution toward higher photon fluxes, and correspondingly
higher values of ionization parameter and density.  \ion{Ca}{2}, since
it is so much more sensitive to depletion compared with the other
elements, constrains the depletion parameter.  And a low dust-to-gas
ratio is required in order to accumulate a large enough column density
to  produce a sufficient column of low-ionization lines without
exceeding our stopping criterion $A_V=2$.   

\subsection{Constraints on Outflow Properties\label{kinematic}}

We compute the kinematic and other properties of the 35 $MFOM=0$ 
solutions, using equations 9 and 11 in \citet{dunn10}.  The results
are listed in the first pair of columns of Table~\ref{table3}. The
radius of the outflow for the $MFOM=0$ solutions lies between 62 and
147 parsecs, with a characteristic value of 110 parsecs.  The outflow
location is strongly  anticorrelated with  $\log(n)-\log(U)$, i.e.,
the photon flux, underlying the strong dependence on this parameter,
as discussed \S\ref{sims}. The log hydrogen column density $[\rm\,
  cm^{-2}]$ is between 22.6 and 22.9, and is strongly anticorrelated
with the dust-to-gas ratio. That is, a large column density is needed
to reach the stopping criterion $A_V=2$ if the dust-to-gas ratio is
low; at the same time, high dust-to-gas ratios are ruled out by the
necessity of having sufficient photoionizing photons to produce the
observed \ion{He}{1}*.   

We used $4,500\rm \, km\, s^{-1}$ as a characteristic outflow
velocity.  It is not obvious what value of the global covering
fraction we should use.  On one hand, the reddening scheme discussed in
\S\ref{goobar} implicitly requires a large covering fraction
approaching $\Omega = 1$.  On the other hand, it is not clear that the
interaction between the BAL wind and starburst gas, as outlined in
\S\ref{picture} occurs along all lines of sight.  So for comparison
with other results, we assumed the rather standard value of the global
covering fraction $\Omega=0.2$, based roughly on the fraction of
quasars observed to have broad absorption lines
\citep[e.g.,][]{dunn10}. 

We obtain very large values of mass fluxes between $600$ and  $2830\rm
\, M_\odot\, yr^{-1}$ compared with other BAL quasars with detailed
kinematic analysis \citep[e.g., Table 10][]{dunn10}.   Likewise, we
also obtain  very large values for the kinetic luminosity, with
$\log \dot E_k$ between 45.6 and 46.3. These large values of kinetic
luminosity lead to very large values of the ratio of the kinetic
luminosity to the bolometric luminosity, between 0.47 and 2.2.  While
such large values imply that the outflow system in Mrk~231 is capable
of vigorous  feedback, it is not clear whether acceleration mechanisms
could produce such large values, especially values greater than 1.  

These extremely large values of mass flux, kinetic luminosities, and
ratio of kinetic to bolometric luminosities occur principally because
of the very large column densities necessary to produce the
\ion{Na}{1} absorption.  The radius is not exceptionally large
compared with other quasars that have apparently low density outflows
in which kiloparsec-scale outflows can be found \citep[e.g., Table 10
  in ][]{dunn10}.  The column density is, however, two to three orders
of magnitude larger than typical column densities found for the
kpc-scale outflow \citep[e.g., 19.9--20.8, Table 10,][]{dunn10}.  The
kinetic luminosity and mass flux depend linearly on column density, so
for those objects, the log kinetic luminosity ranges from 43.7 to
45.7.  { More recently, \citet{arav13} inferred a kinetic luminosity of
$10^{45}\rm \, erg\, s^{-1}$ for the quasar HE~0238$-$1904. However,
  their column density estimates were still rather low
  ($19.8$--$20.9$), and the ratio of the kinetic luminosity to the
  bolometric luminosity was only 1\% for that very luminous quasar.  }

\subsection{[\ion{O}{3}] Line Emission\label{oiii}}

The density of the illuminated side of the gas slab in our simulations
is low, similar to densities present in the narrow-line
region of AGN.   These low densities yield $MFOM=0$ because
the solutions tend to lie along constant values of ionizing flux.  High
densities correspond to low values of ionization parameter, which do
not produce sufficient \ion{He}{1}*.    At any rate, for the favored low
values of density, the \ion{H}{2} region of the outflow, at least,
can be expected to emit lines characteristic of the NLR.  For reference,
the critical density for [\ion{O}{3}]$\lambda 5007$, one of the
brightest NLR lines, is $6.8\times 10^5\rm \, cm^{-3}$ \citep{of06}.
Even in pressure balance, there is a considerable amount of gas in our
simulations with densities smaller than that value.  Thus, our
outflows can be expected to produce line emission.   { We emphasize
  that this putative [\ion{O}{3}] emission should be distinguished
  from NLR [\ion{O}{3}] emission, which is often observed to be relatively
  weak in BALQSOs \citep[e.g.,][]{boroson02}, and is not observed at
  all in Mrk~231.}

We extracted the [\ion{O}{3}] fluxes from the {\it Cloudy} simulations,
and then computed the equivalent width with respect to the {\it Cloudy}
transmitted continuum.  This continuum is an approximation for our 
physical situation, as it is computed using slab reddening, where as, as
we have shown in \S\ref{goobar}, the extinction in Mrk~231 is more
appropriately circumstellar.  However, since the result is not subtle,
the transmitted continuum served our purpose. Assuming again a global
covering fraction $\Omega = 0.2$, we found a range of equivalent widths
between 146 and 270\AA\/.  This would be a very strong line that is
clearly not seen in the spectrum. 

There are several ways in which the predicted [\ion{O}{3}]
line could be hidden.  First, the velocity of the absorption line is
the radial component of the velocity along our line of sight.  The
global outflow, if 
present, would be expected to be directed in a range of directions,
yielding a range of radial velocity components.  That would smear the
line,  decreasing its contrast with the continuum.  Second, it is not
clear that we would be able to see all of the line emission.  That is,
emission in the receding wind on the far side of the object may be
attenuated by dust.  So even if  $\Omega=0.2$, we may be able
to see only a small fraction of that.  Finally, it is possible that
the global covering fraction is much less than $\Omega = 0.2$.  Mrk~231
is not a typical BALQSO, and thus its geometry and physical conditions
cannot be expected to conform with the expectations of the general
population.   Nevertheless, the possibility of forbidden line emission 
from low-density ($n \approx 10^4\rm \, cm^{-3}$) BAL outflows is 
interesting, and possibly constraining, and to our knowledge,
previously unexplored.  

\subsection{Properties of the Characteristic Solution\label{char_sol}}

The modified figure of merit identifies a region of parameter space in
which the upper and lower limits and parameter bounds for the four
diagnostic ions are met.  As discussed above, we identified a
characteristic solution among the set of solutions that satisfy
$MFOM=0$.  That solution is characterized by $\log U=-0.5$, $\log
n=3.75$, dust-to-gas ratio parameter equal to 0.075, and depletion
parameter equal to 0.6.  At this point, the absorbing gas lies $110
\rm \, pc$ from the nucleus, and the { inner edge of the} torus,
characterized by $R_{\tau K}=0.31\rm \, pc$, would subtend an angular
diameter of about 19 arcminutes.  As shown in Table~\ref{table3}, the
log outflow kinetic luminosity would be 46.0, and the ratio of kinetic
to bolometric luminosity would be 1.1.

\begin{figure}[!h]
\epsscale{1.0}
\includegraphics[width=6.5in]{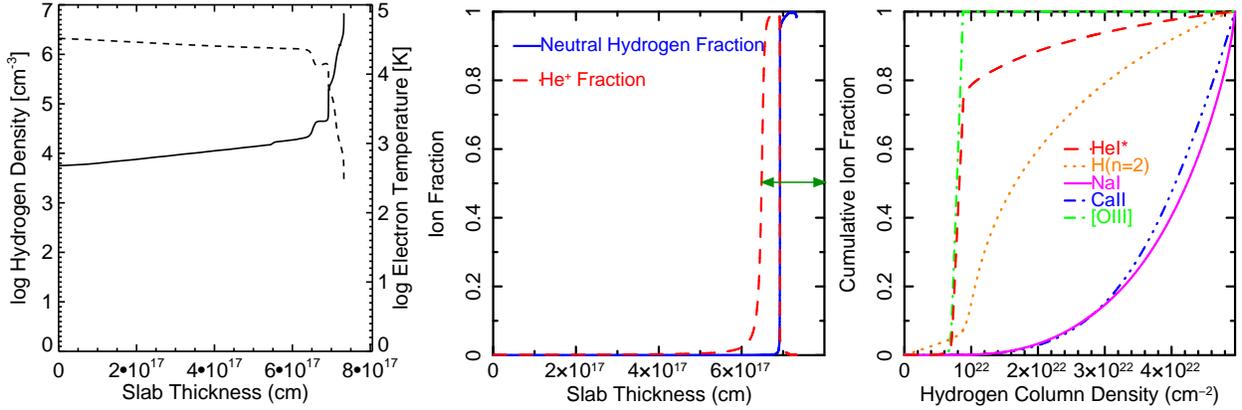} 
 \caption{\small Physical properties of the characteristic solution,
   specified    by $\log U=-0.5$, $\log n=3.75$, dust-to-gas ratio
   parameter equal to    0.075, and depletion parameter equal to 0.6.
   {\it Left:}  The    variation of the hydrogen density (solid line)
   and the electron    temperature (dashed line), as a function of depth in the
   illuminated slab.  In this constant-pressure simulation, as the
   temperature drops, the density increases.  {\it Middle:}  The
   location of the absorbing ions. Metastable helium, which produces
   \ion{He}{1}*$\lambda, \lambda 3889, 10830$ is produced
   principally by  recombination onto He$^+$.  The \ion{Na}{1}D, 
   \ion{Ca}{2}, and undetected Balmer lines are produced in the
   partially-ionized zone, coincident with the region beyond the
   Stromgren sphere and dominated by neutral hydrogen.  Thus, while
   the full slab has a thickness of  $0.24\rm \, pc$, the region of
   the slab where observed lines are formed  { (marked in green)} has a
   thickness of only    $0.027\rm \, pc$.    {\it Right:} The
   cumulative, normalized ion 
   fraction as a function of hydrogen column further illustrates that
   the \ion{He}{1}* absorption originates in  different   gas than the
   low ionization absorption \ion{Na}{1}D and   \ion{Ca}{2}.
   $\lbrack$\ion{O}{3}$\rbrack$ (\S\ref{oiii}) cumulative emissivity
   is   also shown. \label{fig12}}    
\end{figure}

Plots of interesting parameters for the characteristic solution are
shown in Fig.~\ref{fig12}.   The left panel shows the temperature and
density as a function of depth through the slab.  As the photoionizing
continuum traverses the slab, electrons are liberated, heating the
gas. Eventually, the continuum runs out of photons able to ionize
hydrogen, the number of photo-electrons decreases, and the gas becomes
much cooler, especially beyond the hydrogen ionization front, located,
for this simulation, at $6.9\times 10^{17}\rm \, cm$ from the
illuminated side of the slab.  In this constant pressure simulation,
the drop in temperature must be compensated by an increase in density.
Thus, the  gas at the back of the slab has, in effect, a much lower
ionization than it would in a constant density simulation.

The central panel of Fig.~\ref{fig12} shows the location of the
optical and infrared absorbing ions in the illuminated slab.
\ion{He}{1}* is formed by recombination onto He$^+$, so it is
coincident with that ion.  The low-ionization lines \ion{Ca}{2} and
\ion{Na}{1} instead originate in the partially-ionized zone where
hydrogen is predominately neutral.  Although the full slab has a
thickness of   $0.24\rm \, pc$, the region where the lines discussed
in this paper are formed has a thickness of only $0.027\rm \, pc$. 
Note that other higher-ionization lines, e.g., \ion{N}{5}, would be
formed at shallower depths.

The right panel shows the cumulative ion fraction and [\ion{O}{3}]
emission as a function of the hydrogen column density, further
illustrating the location difference in the line-forming regions among
the ions.  It also illustrates that, because of the density gradient,
most of the gas is located in the region where the optical and
infrared lines are formed.   The \ion{He}{1}* nearly saturates
abruptly at rather small column densities; interestingly, some
\ion{He}{1}* (and likewise He$^+$) is present beyond the hydrogen
ionization front, which is located in this simulation at a hydrogen
column density of $8.8\times 10^{21}\rm \, cm^{-2}$.  We don't
understand why this happens, since in many AGN photoionization
scenarios, the continuum runs out of helium-ionizing photons
(ionization potential equal to $24.6\rm \, eV$) before it runs out of
hydrogen ionizing photons (ionization potential equal to $13.6 \rm \,
eV$).   

\subsection{Other Absorption Lines\label{other}}

In this paper, we discuss only the absorption lines observed in the
optical and infrared.  Other absorption lines, including those from
\ion{Fe}{2}, \ion{Mg}{2}, and \ion{Mg}{1}, have been reported by
\citet{smith95}, while a broad \ion{C}{4} line was reported by
\citet{gallagher02}. In principle, better constraints on our
photoionization model could be obtained by combining the near-UV line
measurements with the optical and infrared line measurements.
However, both UV spectra have very poor signal-to-noise ratios, making
constraint of the continuum and measurement of the absorption  lines
quite difficult.  The second, possibly more important issue is that
the partial covering may be wavelength dependent, making absorption
lines appear shallower than they really are.    There is no reliable
way to estimate the covering fraction with the data at hand if it is
wavelength dependent. Newly approved {\it HST} COS and STIS
observations may provide much better constraints than the archival
data.     

Nevertheless, we discuss one particular line to illustrate the
potential problems. \citet{smith95} show a clear detection of
\ion{Mg}{1}$\lambda 2852$ in their FOS spectrum.  But the line does
not appear to be very broad or deep, at least compared with the
\ion{Na}{1}D line. \citet{smith95} does not provide an estimated limit
on \ion{Mg}{1} column density, so we estimate it from the reported
equivalent width and the digitized spectrum, finding a lower limit on
the log column density of $\sim 11.3$, much smaller than the measured
\ion{Na}{1}D column.  However, our {\it Cloudy} models
(Table~\ref{table3}) predict a range of \ion{Mg}{1} log column
densities between 15.0 and 15.4, similar to the \ion{Na}{1} column
densities.  One possible explanation for this 
discrepancy is that the covering fraction of the nuclear starburst is
higher in the NUV than it is for the continuum source.  (This is
conceivable, for example, if the starburst is the origin of the
circumstellar dust).  That could make the \ion{Mg}{1} absorption line
appear much shallower than the optical and IR lines, leading to a low 
estimate of the column density. Mrk~231 is close enough that the
nuclear starburst can be resolved, and so it may be possible to probe
differential reddening/covering for the nuclear emission versus the
starburst emission with the upcoming 1-orbit {\it   HST} STIS
observation.   

Likewise, the {\it Cloudy} models predict other common UV BAL lines to
have high column densities.  For example, \ion{C}{4} is predicted to
have a log column density of between 17.2 and 17.4.  This line would
be saturated.  The only line predicted to be somewhat weak is
``\ion{Fe}{2} High'', as we refer to the transitions between 2692 and
2773 \AA\/ that arise from levels between 0.98 and 1.1 eV above the
ground state \citep{lucy14}.  Those excited-state \ion{Fe}{2} atoms
are predicted to have a log column density between 13.2 and 14.5, while
the transitions near the ground state (``\ion{Fe}{2} Low'') are
predicted to have log column density of 16.7--17.0.  This happens 
because many of the excited-state \ion{Fe}{2} transitions have
relatively high critical densities \citep[e.g., higher than $10^5\rm
  \, cm^{-3}$,][]{korista08}.  Our gas has too low density to promote
many of the iron ions to the excited states.  Depending on the width of
the absorption lines, it is then possible that \ion{Fe}{2} High would
not be saturated.

\section{Beyond the Initial Model\label{beyond}}

In \S\ref{sims}, we showed that we could obtain a family of solutions
that produced the ionic column densities that we measured and are
consistent with  upper and lower limits.  However, we needed to make a
series of assumptions to construct this model.  In this section, we
investigate the consequences of these assumptions, one by one, to
determine how they influence the location, energetics, and other
properties of the absorber. 

\subsection{Dust in {\it Cloudy}\label{dust}}

The current version of {\it Cloudy} is quite sophisticated  and
flexible in its handling of dust.  Several dust models are built in,
and a user may even input his or her own dust model.  As noted above,
we used the built-in ISM dust for our simulations, since Milky Way
dust was favored in our reddening analysis (\S\ref{goobar}).  It may 
be that the dust in Mrk 231 has much different absorption and 
scattering properties than Milky Way dust, given that it is subject
both to the quasar continuum and the nuclear starburst.
Qualitatively, for example, a different type of dust may produce a
smaller or larger extinction in the helium continuum.  That would
increase or decrease our inferred dust-to-gas ratio, to meet the
requirement that sufficient photons remain in the He$^+$ continuum to
produce the observed \ion{He}{1}* column density. 

Despite this flexibility, the  {\it Cloudy} user cannot specify
  variable dust properties (i.e., variable dust-to-gas ratio) as a   
function of depth of a slab.  As discussed in \S\ref{picture}, based
on the kinematic differences between the \ion{He}{1}* lines and the
low-ionization lines \ion{Ca}{2} and \ion{Na}{1}, we suggested that the
\ion{H}{2} region of the outflow is a traditional BAL wind, and is
therefore relatively free of dust.  But dust must be present in the
\ion{H}{1} region of the  outflow, since depletion is necessary to
explain the \ion{Ca}{2} column density.  If the \ion{H}{2} region were
dust free, more \ion{He}{1}* would be produced  since, as discussed in
\S\ref{sims}, when dust is present, \ion{He}{1}* is suppressed for
high photon fluxes as the \ion{He}{1} continuum is depressed by
extinction.  

Is dust necessary in the {\it Cloudy} simulation?  That is,
perhaps all of the absorption occurs interior to the circumstellar
scattering region, in which case the absorber may be free of dust.
To test this possibility, we ran several dust-free models, starting at
the parameter values of the characteristic solution, but using a range
of hydrogen column densities as the stopping criteria.  We find that
a log hydrogen column density of 23.7 is necessary to produce the lower
limit on the \ion{Na}{1}.  This an order of magnitude larger than the
column density inferred from the characteristic solution.  This model
also produced insufficient calcium, although that could be addressed
by adjusting the depletion.  Regardless, an order-of-magnitude larger
column density would predict an order of magnitude larger kinetic
luminosity, making an inferred kinetic luminosity many times larger
than the bolometric luminosity.

Another issue is that while we used circumstellar dust to explain the
reddening, the {\it Cloudy} models use slab-type dust.  It is clear
that the circumstellar reddening cannot be present throughout the
absorbing gas because the extremely strong blue extinction would
remove all the He$^+$ continuum  photons, and insufficient
\ion{He}{1}* would result.  However, we may imagine again that the
\ion{H}{2} region, where the \ion{He}{1}* absorption is produced, lies
inward to the scattering region, and the partially-ionized zone lies
in gas in which the circumstellar reddening is appropriate.  Due to
strong blue opacity, the partially-ionized zone may have a very low
skin-depth, transitioning very quickly to gas in which neutral sodium
and Ca$^+$ can exist.  Then, it is possible that the required
absorption lines could be produced with smaller column density, and
perhaps no density increase would be required.  We cannot simulate
this situation with the tools at hand, however.    

\subsection{The Constant Pressure Assumption\label{const_press}} 

In \S\ref{sims}, we assumed constant pressure in our {\it Cloudy}
model in order to provide a density gradient.  However, constant
pressure is not likely to be physically realistic.  First, the outflow 
velocity is much greater than the speed of sound (approximately $10\rm
\, km\, s^{-1}$ in the \ion{H}{2} region) so it is not clear how
pressure equilibrium would be attained in the dynamical environment.
Second, as discussed in \S\ref{abs_lines}, there is evidence from the
line profiles that the inner gas is moving radially outward with a
larger velocity than the outer gas, and we suggested (\S\ref{picture})
a scenario in which the inner gas from an quasar BAL wind impacts and
scoops up processed gas from the starburst.  In this case, a shock may
be present; alternatively, the processed gas may have higher density
than the BAL wind.  

It turns out, however, that we do not need to assume constant
pressure to obtain the measured ionic column densities; all we need is
a density increase between the \ion{He}{1}* 
absorbing gas, and the \ion{Na}{1} absorbing gas.  To test 
how much of a density increase is necessary, we ran a series of
{\it   Cloudy} models using density laws, i.e., we specified density as
a function of depth from the illuminated side of the gas slab. To
build the parameterized density 
laws, we started with the characteristic solution, assuming constant
pressure, and extracted the density as a function of depth.   Since
\ion{He}{1}* is produced in the \ion{H}{2} region and the low
ionization lines are produced in the partially-ionized zone, we placed 
the density increase at the hydrogen ionization front.  We retained
the density profile in the \ion{H}{2} region, since we know that it 
provides the required amount of \ion{He}{1}*. We modified the density
in the partially-ionized zone,  assuming a constant value throughout,
for a range of multiplicative enhancement factors relative to the
density in the region where He$^+$ is found (for example, for the
characteristic solution, $\log(n)=4.66$ at this point).  This density
profile is ad hoc, although  one might imagine that an approximate
step function density profile could be obtained in a shock.   

For no or little density enhancement, the sodium column is low, too
low to produce a detectable \ion{Na}{1} absorption line.  We find that
an enhancement by a multiplicative enhancement factor of 14.5
(i.e., increase to a log hydrogen density of 5.8) will yield required 
\ion{Na}{1} column.  So, although the constant pressure solution 
illustrated in Fig.~\ref{fig12} indicates a density jump 
of $\sim 1000$ from the illuminated side to the back end, in fact, we
only really need a factor of $14.5$ to explain the \ion{Na}{1} column
density. 

\subsection{Elevated Abundances\label{abund}} 

If the swept-up gas originates in the nuclear starburst, the
abundances may be modified by the stellar processing.  AGB
stars can eject gas with large enhancements in abundances
\citep[e.g.,][]{kraft94}, depending on the length of time in certain
evolutionary states.  For example, in the most favorable case, sodium
can be enhanced by a factor of 100. 

We computed an altered abundance pattern by assuming that the
starburst rapidly used up all the gas in the central regions of the
Mrk 231, and that the gas contained in the AGN outflow was entirely
the product of enriched mass loss from those stars that became Type II
supernovae (SN) or AGB stars before some designated time.  We took AGB
yields from \citet{karakas10} and yields for more massive stars from
\citet{cl04}, and we linearly interpolated the yields for the various
elements as a function of stellar mass between the two studies.  In
all cases we used the yields from solar-abundance models.  The
computation assumed a \citet{sal55} mass function with $x = -2.35$,
and we took the age since the starburst as 100 Myr
(cf.\ Fig~\ref{fig4}); this sets the lower mass limit in the
computations to $4 M_\odot$, which is the mass of a star with this
lifetime in the Karakas models. We computed both the combined AGB + SN
case and (for comparison) AGB-only enrichment.  The former computation
used an upper mass bound of $20 M_\odot$, while the upper bound for
the latter case was set at $7 M_\odot$.  The luminosity-function
weighted yields were most sensitive to the upper bound for the
combined case, but did not depend strongly on the lower mass limit.
The resulting abundance enhancements relative to hydrogen are listed
in Table~\ref{table4}.  

We tested the effects of these enhanced abundances on
the characteristic solution.  We assumed that the gas is purely
reprocessed gas from the starburst, i.e., with no un-reprocessed gas
mixed in, and therefore, these numbers represent the maximum
enhancement we could expect.  For the low-mass stellar evolution
scenario, the \ion{He}{1}* column was increased by 0.15 dex and the
\ion{Na}{1} was increased by 0.22 dex.  When the higher mass stars are
included, we found \ion{He}{1}* enhancement by 0.3 dex and \ion{Na}{1}
enhancement by 1.34 dex (i.e., a factor of 22 times larger).  In both
cases, \ion{Ca}{2} and \ion{H}{1} in n=2 remained in observed bounds. 

\begin{deluxetable}{lcccc}
\scriptsize
\tablecaption{Fractional Abundance Enhancements in 100 Myr Starburst}
\tablehead{
\colhead{Isotope} & 
\colhead{AGB\tablenotemark{a}} & 
\colhead{AGB/1H\tablenotemark{b}} & 
\colhead{AGB$+$SN\tablenotemark{c}} & 
\colhead{AGB$+$SN/1H\tablenotemark{d}}}
\startdata
1H & 0.918 & 1.000 & 0.850 & 1.000 \\ 
4He & 1.191 & 1.298 & 1.255 & 1.477 \\ 
12C & 0.851 & 0.927 & 1.873 & 2.203 \\ 
14N & 5.502 & 5.995 & 4.955 & 5.830 \\ 
16O & 0.869 & 0.947 & 1.896 & 2.231 \\ 
20Ne & 1.000 & 1.089 & 5.716 & 6.725 \\ 
22Ne & 2.542 & 2.770 & 7.148 & 8.410 \\ 
23Na & 1.709 & 1.862 & 11.346 & 13.350 \\ 
24Mg & 0.980 & 1.068 & 2.635 & 3.100 \\ 
27Al & 1.061 & 1.156 & 2.961 & 3.484 \\ 
28Si & 1.004 & 1.094 & 6.067 & 7.138 \\ 
32S & 0.996 & 1.086 & 4.436 & 5.219 \\ 
34S & 1.003 & 1.092 & 1.394 & 1.640 \\ 
56Fe & 0.995 & 1.085 & 2.186 & 2.573 \\ 
\enddata
\tablenotetext{a}{Ratio of isotope to solar abundance assuming a mass function
slope of $-2.35$ for a $100\rm \, Myr$ starburst including AGB stars
down to 4 solar masses.}
\tablenotetext{b}{Ratio of isotope to solar abundance normalized to the hydrogen
ratio assuming a mass function slope of $-2.35$ for a $100\rm \, Myr$
starburst including AGB stars down to 4 solar masses.}
\tablenotetext{c}{Ratio of isotope to solar abundance assuming a mass function
slope of $-2.35$ for a $100\rm \, Myr$ starburst including AGB stars
down to 4 solar masses and Type II supernovae.}
\tablenotetext{d}{Ratio of isotope to solar abundance normalized to the hydrogen
ratio assuming a mass function slope of $-2.35$ for a $100\rm \, Myr$
starburst including AGB stars down to 4 solar masses and Type II
supernovae. }
\label{table4}
\end{deluxetable}
\normalsize

The effect of these enhancements is to expand the
allowed region of parameter space.  As discussed in \S\ref{char_sol},
the $MFOM=0$ solutions are found in a narrow band of photoionizing flux
(and therefore radius) because of the opposing constraints provided by
the \ion{He}{1}* and \ion{Na}{1} columns on this parameter.  The
stellar reprocessed gas has enhanced abundances for both helium and
sodium, and thus the column densities for both parameters are
increased.  Since the criterion for both parameters is a lower limit,
the effect is that a broader range of photoionizing fluxes produce
column densities in the acceptable range.   

The second effect is to permit $MFOM$ solutions with a larger value of
dust-to-gas ratio.   The dust-to-gas ratio in the characteristic
solution was constrained to be low so that there were sufficient
photons in the helium continuum to produce the observed \ion{He}{1}*.
A low dust-to-gas ratio also contributes in allowing sufficient column
density to reach the neutral gas in which the \ion{Na}{1} column is
produced.  A larger dust-to-gas ratio would lead to a lower hydrogen
column density, and a lower inferred kinetic luminosity.

\subsection{Continuum Shape Assumption\label{continuum}}

The simulations presented here employ the {\it Cloudy} ``Kirk''
continuum, suggested as a representative AGN continuum by
\citet{korista97}.  This continuum is fairly hard, with an
$\alpha_{ox}$  measured from the continuum of $-1.45$.  Based on our
inferred intrinsic 2500\AA\/ monochromatic luminosity of $1.2\times
10^{30}\rm \, ergs\, s^{-1}\, Hz^{-1}$, and the relationship between
UV luminosity and $\alpha_{ox}$ presented by \citet{just07}, an
$\alpha_{ox}=-1.51$ would be predicted for Mrk~231.  
Evidence has been increasing that at least some BALQSOs have
intrinsically { X-ray} weak spectral energy distributions
\citep[e.g.,][]{sabra01,clavel06,grupe08,luo13}.  Mrk~231 was
recently observed in the X-ray using {\it Suzaku} by
\citet{piconcelli13}, who reported an increase in X-ray flux and
appearance of the primary continuum at $<10\rm \, keV$.  They recorded
an intrinsic 2--10 keV luminosity of $3.3\times 10^{43}\rm \, erg\,
s^{-1}$.  Assuming that the intrinsic spectrum can be described by a
$\Gamma=1.9$ power law, an $\alpha_{ox}\approx -1.7$ is inferred.
{ Recently, a value of $\alpha_{ox} \sim -1.7$ was measured using                                          
  {\it NuSTAR} data \citep{teng14}.  Therefore Mrk~231 is now known 
to be intrinsically X-ray weak. }

As discussed in \citet[][Appendix A]{leighly11}, a higher column
density is generally required to produce a   large \ion{He}{1}* column
when the continuum is soft, simply because   soft continua lack
photons in the He$^+$ continuum  ($> 24\rm \, eV$) that are needed  to
ionize helium.  In other words, a more intense continuum and thicker
column would be required to produce the same \ion{He}{1}* column for a
softer continuum.  For example, for the semi-empirical   spectral
energy distributions introduced in \citet{casebeer06} and   used in
\citet{leighly07}  and \citet{leighly11}, an   $\alpha_{ox}=-1.7$
continuum requires 0.2 dex higher column density,   and a 0.4 dex
higher photon flux to produce the same \ion{He}{1}* column density as
a $\alpha_{ox}=-1.4$ continuum.  From a photoionization model point of
view, a higher column density might  be difficult to muster, given that
in the dusty model, a higher column will incur more reddening, or,
alternatively, since our stopping criterion in the simulations is the
reddening, a lower dust-to-gas ratio.   

However, these results depend on the precise shape of the continuum.
The \citet{casebeer06} SEDs link the UV cutoff with the UV luminosity
through $\alpha_{ox}$  according to $T^{1/4}$ \citep[see ][for
  details]{casebeer06}.  If, for example, the UV emission remained
unchanged, but the coronal X-rays were weak or missing \citep[a 
possibility that has been suggested recently for two BALQSOs based on
recent  {\it NuSTAR} results;][]{luo13} it may still be possible
to  produce strong \ion{He}{1}*.  We tested this possibility by 
performing  {\it Cloudy} modeling using a modified version of
the``Kirk'' AGN {\it Cloudy} to be weaker in the X-rays, created by
simply cutting the spectrum at a  particular energy, and then
decreasing the flux shortward of that  point by a multiplicative
factor.  To maximize the effect, we first varied the energy of the cut
point, and determined the value required to produce most (within 97\%)
of the \ion{He}{1}* obtained using the unmodified spectrum.  That
energy was $\sim 45\rm \, eV$.  We then tried two continua, created by
decreasing the  normalization for $E> 45\rm \, eV$ by factors of 10
and 100.  These  continua are ad hoc, but considering that the shape
of the extreme UV continuum is not known, they served to test the
effect.  We used the parameters of the characteristic solution as the
starting point, but retained the normalization of the spectrum.  The
ionization parameter was then decreased from $-0.5$ to $-0.61$ and
$-0.62$ for the continua  attenuated by factors of 10 and 100,
respectively.      

We found that the  \ion{He}{1}* column density of the modified spectra
actually increased over that of the unmodified spectra. Specifically,
the unmodified spectrum yielded a \ion{He}{1}* log column density of
14.83, but the spectra cut by factors of 10 and 100 yielded \ion{He}{1}*
columns of 15.14 and 15.23, respectively.  Examination of the
simulation results show that these X-ray weak continua shift the
overall ionization of the gas to intermediate- and low-ionization
lines.  A similar effect was seen in simulations of emission lines of
the X-ray weak quasar PHL~1811 \citep{leighly07}.  The sodium column
density, on the other hand, decreased to 14.73 and 14.78,
corresponding to $-0.089$ and $-0.071$ dex, respectively.  These small
changes imply that the factor most important in determining the
\ion{Na}{1} column density { is how much of the sodium is ionized
  to Na$^+$, which is controlled by the strength of the spectrum
  shortward of 2412\AA\/, and is unchanged in these simulations.} 

The increase in the simulated \ion{He}{1}* column means that the
acceptable parameter space would be expanded, specifically toward
lower photon fluxes and therefore larger radii.  To test the
consequences of that shift, we ran a sequence of models while
decreasing the overall continuum normalization.  We found that the
\ion{He}{1}* was reduced to the non-modified-continuum level of the
characteristic solution when the continuum decrease multiplicative
factors were 5.4 and 7.3 for the 10 times X-ray weak and 100 times
X-ray weak continua respectively.  These models produce large column
densities of sodium (larger by factors of 2.8 and 3.7 over the
characteristic solution, respectively), and the other lines were  
consistent with the required limits and ranges.  Also, like the
abundance enhancement simulations, a larger value of dust-to-gas ratio
would be allowed, corresponding to a reduced column density and
kinetic luminosity.  

\section{Optimizing Parameters\label{optimize}}

Our initial model, described in \S\ref{cloudy}, included a number of
assumptions, and also resulted in a very large column density outflow,
and correspondingly large and possibly unphysical values of kinetic
luminosity.  In \S\ref{beyond}, we investigated these assumptions
individually, specifically to see if modifying them would produce an
outflow with lower column density.   We found that a lower hydrogen
column density could probably be attained if the gas abundances were
modified by reprocessing by the starburst, or if the incident
continuum were softer.  In addition, we showed that constant pressure
was not needed to obtain a satisfactory solution; rather, a modest
step increase in density, such as might be produced in a shock, could
also produce required ionic column densities.    

\begin{figure}[!h]
\epsscale{1.0}
\begin{center}
\includegraphics[width=3.5truein]{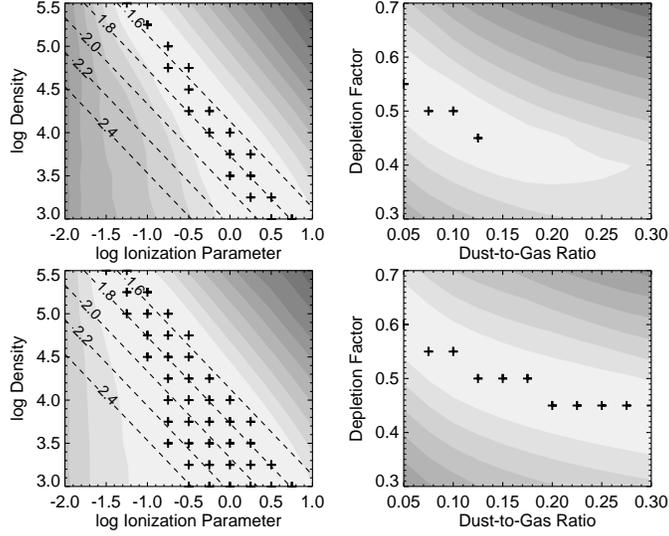}
\end{center}
\caption{\small Contours of the modified figure of merit (MFOM) as a
  function of simulation parameters, for a fixed $A_V=2$, in
  orthogonal directions to the characteristic solution
  (Table~\ref{table3}).    These contours show the case
  described in \S   \ref{optimize}, wherein the abundances were
  enhanced, the SED   softened, and a density step was implemented,
  rather than constant   pressure assumed.   Note that to facilitate
  comparison with Fig.~\ref{fig7}, the ionization parameter refers to
  the original continuum; the actual ionization parameters for the
  soft continuum are 0.11 dex lower.  The plus
   signs show where $MFOM=0$, i.e., where the solutions are consistent
   with the limits and bounds (\S\ref{sims}).  The dashed lines show
   contours of constant log distance of the absorber from the central
   engine, in  units of parsecs.  {\it Top:} The contours for a
   log density enhancement of 0.6.  This panel shows that a solution
   can be found for a very modest density jump (factor of 4), but the
   parameters are very highly constrained.    
 {\it Bottom:} The contours for a
   log density enhancement of 1.4. This panel shows that for a large
   density enhancement (factor of 25), parameter space opens up
   considerably.    \label{fig9}} 
\end{figure}

In this section, we close the loop by running a full grid of models
with a softer continuum and enhanced abundances, and using two density
step-function enhancement factors.  We then extracted a characteristic
solution and range for these optimized parameter choices for each of
the density step functions.  Specifically, we used the soft continuum
with the flux density decreased by a factor of 10 at energies $>45\rm
eV$, and abundances modified by the medium-mass stars only.  We test
two density enhancement factors: 0.6 and 1.4.  Thus, the density in
the partially ionized zone increased by a factor of $10^{0.6} \approx
4$ and $10^{1.4} \approx 25$ with respect to the density in the He$^+$
zone at the hydrogen ionization front.  The results are given in
Table~\ref{table3} and Figure~\ref{fig9}.

In some respects, the solutions are very similar to our original
solution.  Specifically, the characteristic solution, i.e., the
solution lying the minimum distance from all other $MFOM=0$ solutions,
is very nearly the same in all three cases.  The larger difference
lies in the ranges of parameters.  In many cases, the range of
parameters is  larger for the density-jump simulations compared with
the constant pressure simulation.  This illustrates the fact that the
softer continuum and enhanced abundances widens parameter space.  Most 
significantly, the range of parameters, especially in the case of the
larger density jump, encompass outflow kinetic energies that are
arguably more reasonable (i.e., less than the bolometric luminosity)
than those obtained in \S\ref{kinematic}.  At the same time, the
characteristic solutions are still all located  $\sim 100\rm \, pc$
from the central engine.

The two density-jump solutions are distinctly different, however; this
is best illustrated by Fig.~\ref{fig9}.  For a density jump by a 
factor of 4, the solutions are highly constrained in a narrow region
of parameter space.  These solutions
require a small dust-to-gas ratio and correspondingly large column 
density, because the small density jump only modestly lowers the
ionization parameter in the partially-ionized zone, and a large column
density is therefore required to obtain the observed \ion{Na}{1}
column.  In contrast, for a density jump by a factor of 25, parameter
space is very large, the dust to gas ratio can be as large as 0.3 (the
upper limit in the simulation grid), so the column density can be low.
This is because the large density enhancement produces a low
ionization parameter in the partially-ionized zone, and sufficient
\ion{Na}{1} can be produced at low column density.   

All of the simulations presented in this paper lead us to conclude
that a density gradient or jump  is necessary to produce the observed 
\ion{He}{1}* and the \ion{Na}{1} ionic column densities. A larger
question is, how much of a density jump could be produced in nature? If
our surmise is correct, and the absorption lines are formed in the
interaction region between the normal quasar outflow and the
starburst, then in principle, the equations of motion could be written
down and solved.  In that case, the density jump incurred in the shock
could be obtained. 

We note that in the case of a supernova interacting with the
circumstellar medium, the outflow velocities are similar and a high
density gradient can be achieved. At minimum, we expect that the
density contrast will be given by the Rankine-Hugoniot conditions and
that for a gas with a ratio of heat capacities $\gamma = 5/3$, the
density contrast would be $\frac{\gamma+1}{\gamma-1} = 4$. However,
taking the analogy with supernovae further, higher density contrasts
are possible. \citet{CF89} studied the density contrast obtained as a
function of the density profile in the outflow. They assumed that the
outflow density profile followed a powerlaw $\rho \propto r^{-n}$ and
that the circumstellar medium had a density profile of a constant
velocity wind, $\rho \propto r^{-2}$. In this case they found that for
$n = 20$, density contrasts can exceed $150$. In fact, for very steep
ejecta profiles $n = 50$, density contrasts can exceed $1000$
\citep{CF94}. The steep density outflow profiles in supernovae are the
natural result of the supernova shock running down the original steep
profile that was established by hydrostatic equilibrium in the
progenitor star, whereas in the case of an AGN the outflow will likely
be accelerated. The material that the outflow runs into, however,
should have a density profile not too far away from the $r^{-2}$ of a
constant velocity wind. Thus, it may not be too far afield to suggest
that large density contrasts could occur.

\section{Summary and Future Directions\label{summary}}

We present the first J-band infrared, and new blue optical spectra of
Mrk 231.  Combining these with spectra taken from the literature, we
discovered a physical solution for the unusual reddening in this object.
{\it Cloudy} modeling revealed unusual physical conditions required to
produce both \ion{He}{1}* and \ion{Na}{1} absorption lines in the same
gas, and inspired a physical interpretation involving an interaction
of a quasar BAL wind with gas ejected from a nuclear starburst.  The
specific results are as follows:

\begin{itemize}
\item We fit the broad band spectrum from $\sim 2300$ \AA\/ to $2.3\,\rm
  \mu m$ with a continuum model, a low-temperature black body, and a
circumstellar reddening and extinction curve originally proposed to
explain the low values of observed $R_V$ in SNe Ia \citep{goobar08}.
Circumstellar reddening is distinguished by large optical depths,
approaching 1, producing increased extinction in the blue \& UV (due
to longer scattering path lengths), along with light scattered back
into the line of sight as a secondary effect.  We obtained an excellent
fit using extinction-shape parameters similar to those for Milky Way
dust (\S\ref{goobar}).  We suggested that the dust is produced in the
nuclear starburst lying $\sim 100\rm \, pc$ from the nucleus
\citep[e.g.,][]{davies04}.    

\item The infrared spectrum revealed a deep \ion{He}{1}* absorption
  line that is very similar in profile to  the well-known \ion{Na}{1}
  absorption line.  Two newer infrared spectra revealed evidence
  for the   appearance of a new absorption line component near
  $11,000 \rm   \,   km\, s^{-1}$.    We modeled the optical and 
  infrared spectra to   extract the line profiles (\S\ref{nuclear},
  \S\ref{hei10830}).   After accounting for the $\sim 100\rm \, Myr$
  nuclear starburst   contribution, we inferred that the absorber
  essentially completely   covers the quasar continuum emission
  region.  The absorption line   profiles indicated that the lines
  produced in the \ion{H}{2} region,   including \ion{He}{1}*$\lambda
  10830$ and \ion{He}{1}$\lambda   3889$, have a higher velocity than
  the low-ionization lines produced   in the partially-ionized and
  neutral gas, including \ion{Ca}{2} and   \ion{Na}{1} (\S\ref{abs_lines_2}).  

\item {\it Cloudy} modeling showed that in order to produce both the
  \ion{He}{1}* and the  \ion{Na}{1} absorption lines, a density
  increase is required between the  \ion{H}{2} region, which produces
  the \ion{He}{1}* lines, and the partially-ionized/neutral region,
  the origin of the \ion{Ca}{2} and \ion{Na}{1} lines.  We first
  modeled   this effect as a constant pressure gas (\S\ref{cloudy}).
  The models   are able to  produce the measured column densities and
  limits if the gas lies   $\sim 100\rm \, pc$ from the central
  engine, i.e., in the vicinity   of the nuclear starburst
  \citep[e.\ g.,][]{davies04}.  These facts,   along with the velocity
  differences of the lines, and the inferred   full covering, led us 
  to a physical scenario in which  the   \ion{He}{1}* absorption
  arises in a quasar BAL outflow that  impacts   and compresses dusty
  gas originating in the starburst, and   this swept up gas is the
  origin of the \ion{Na}{1} and \ion{Ca}{2}   lines (\S\ref{picture}).
  In addition, we noted that just such an   interaction between a
  quasar outflow and surrounding star-forming   gas may be an example
  of quasar feedback, thought to be necessary to   shut down star
  formation during the co-evolution of black holes and   quasars.

\item In \S\ref{beyond} we examined the effects of modifying the
  assumptions made in our initial {\it Cloudy} model.  In addition,
  the outflow masses and kinetic luminosities inferred were very
  large if the global covering fraction is $\Omega=0.2$.   Constant
  pressure gas may not be a reasonable assumption for a $\sim
  -4,500\rm \, km\, s^{-1}$  outflow, and therefore we instead
  experimented with a density increase between the \ion{H}{2} region and
  the partially-ionized zone that might be physically realized in a
  shock, and found that a density increase works as well as constant
  pressure.  We discovered that the area of parameter space producing
  models    consistent with the measurements is broadened when we
  consider   abundances enhanced by stellar processing in the
  starburst, and when the  spectral energy distribution is soft.  We
  also   suggested that  the  \ion{H}{2} BAL outflow is free of dust,
  while   the  partially-ionized  zone may have circumstellar dust.
  This   situation,  which  could  further reduce the inferred column 
  density and therefore kinetic luminosity, cannot  presently  be
  modeled  using {\it     Cloudy}   (which cannot model variable  dust
  properties as a function of depth).  

\item Finally, in \S\ref{optimize}, we presented full grids using the
  optimal parameters obtained in \S\ref{beyond}.  For a soft spectral
  energy distribution and enhanced abundances, a   density 
  enhancement by a factor of only  four can still produce the lines we
  observe, although over a limited region of parameter space.  A
  density enhancement by a factor of 25 opened up parameter space
  considerably.  A shock would be expected to produce a density
  contrast of 4 for gas with $\gamma=5/3$, and much larger density
  contrasts are inferred in supernovae, depending on the density
  profiles of the outflow and circumnuclear gas.

\item Most of the simulations producing the observed ionic column
  densities favor low densities, i.e., $\log n \approx 4$.  This
  density is characteristic of the narrow-line region in AGN.  We find
  that, assuming a global covering fraction of $\Omega = 0.2$, the
  inferred equivalent width of the [\ion{O}{3}] emission line should
  be 100--200\AA\/.  Such a huge line is not seen in Mrk~231.  We note 
  that the line may not necessarily be predicted to be sharp, but
  rather could be smeared by a range of line-of-sight velocities, or
  obscured by dust.  
\end{itemize}

While we consider the question of the cause of the anomalous reddening
essentially solved in this paper, a number of questions regarding the
absorption lines remain.  Some may be addressed by the upcoming {\it HST}
observations of Mrk~231. For example, additional UV and near-UV
absorption line measurements may be able to further refine the {\it
  Cloudy} modeling; this may be complicated by differential continuum
and starburst covering fractions and extinctions.   Dynamical modeling
may be able to determine whether shocks could produce  density
jumps as large as the simulations require.  Further development of
{\it Cloudy} may ultimately allow specification of dust properties as
a function of depth into the gas slab.  In addition, our models are
1-D, and so cannot explicitly take into account the effect of the
circumstellar reddening on the photoionization results.   At any rate,
the unusual set of circumstances required to produce the observed
optical and infrared absorption lines may explain why \ion{Na}{1}
lines are so rare, or at least why the line in Mrk~231 is so
exceptional; without the interaction of the BAL  wind and starburst
gas along our line of sight, Mrk~231 might instead look like an
ordinary FeLoBAL quasar.

\acknowledgements

KML acknowledges very useful conversations with Dick Henry, Kieran
Mullen, and Angela Speck, and useful comments from the OU Astro
Journal Club.  KML 
thanks Lucimara Martins for providing her  starburst infrared spectral
templates.  KML gratefully acknowledges John Wisniewski's donation of
APO time to the OU astronomy group, and thanks him for taking the
April 7 2014 observations.   KML and DMT thank OU Nielsen Hall system
administrator Andy Feldt for copious help and advice in using the NHN
Condor queue.  KML and ABL acknowledge support through NSF
AST-0707703.  The authors acknowledge Sara Barber's participation in
the KPNO  and IRTF observing runs.  TIFKAM was funded by The Ohio State
University, the MDM consortium, MIT, and NSF grant AST-9605012. The
HAWAII-IR array upgrade for TIFKAM was funded by NSF Grant AST-0079523
to Dartmouth College.

{\it Facilities:}  \facility{Mayall (R-C CCD Spectrograph)},
\facility{IRTF (SpeX)}, \facility{Hiltner}, \facility{ARC: 3.5m
  (TripleSpec)} 

\bibliographystyle{apj}
\bibliography{leighly}

\end{document}